\documentstyle[seceq,preprint,epsf]{jpsj} 

\def\runtitle{Magnetic and Metal-Insulator Transitions  in Two-Dimensional 
Hubbard Models} 
\def\runauthor{Tsuyoshi {\sc Kashima} and Masatoshi {\sc Imada}}  
\title { Magnetic and Metal-Insulator Transitions\\ through Bandwidth Control\\ 
in Two-Dimensional Hubbard Models\\ with Nearest and Next-Nearest Neighbor Transfers  }  
\author {  Tsuyoshi {\sc Kashima} and Masatoshi {\sc Imada} }  
\inst { Institute for Solid State Physics,\\ University of Tokyo,5-1-5 Kashiwanoha,\\ 
Kashiwa, Chiba, 277-8581 }  
\recdate { \today }  
\abst { Numerical studies on Mott transitions caused by the control of 
the ratio between bandwidth and electron-electron interaction ($U$) are reported.   
By using the recently proposed path-integral renormalization group(PIRG) algorithm, physical properties near the transitions in the ground state of two-dimensional 
half-filled models with the nearest and the next-nearest neighbor transfers 
($-t$ and $t'$, respectively) are studied as a prototype of geometrically frustrated system.
The nature of the bandwidth-control transitions shows sharp contrast
with that of the filling-control transitions:  First, the metal-insulator and magnetic transitions are separated each other 
and the metal-insulator (MI) transition occurs at smaller $U$, although the both 
transition interactions $U$ increase with increasing $t'$.  Both transitions do 
not contradict the first-order transitions for smaller $t'/t$ while the MI 
transitions become continuous type accompanied by emergence of {\it unusual 
metallic phase} near the transition  for large $t'/t$.  A nonmagnetic insulator phase is stabilized between MI and AF transitions.  The region of the nonmagnetic insulator becomes wider with increasing $t'/t$.  The phase diagram naturally connects two qualitatively different limits, namely the Hartree-Fock results at small $t'/t$ and speculations in the strong coupling Heisenberg limit.  
}  
\kword { quantum simulation, strongly correlated electron, Hubbard model, 
metal-insulator transition, magnetic transition, Mott transition,  
quantum phase transition, spin liquid, J$_1$-J$_2$ model, geometrical frustration }  
\begin{document} 
\sloppy \maketitle  
\section{Introduction} 
\label{INTRO} 
The Mott transition between metallic and insulator phases has been a  subject of general interest since its proposal in the middle of the last  century.  The original issue was related with the failure of the band theory in explaining the Mott insulators such as NiO.  According to the band theory, the highest filled band is completely filled for insulators and is partially filled for metals.  The Mott insulators, however, have partially filled bands and also have the insulating nature.  To understand the insulating nature, Mott introduced the concept of electron localization caused by the mutual correlation effects~\cite{MIT,Mott}.
In strongly correlated electron systems, the interplay of strong quantum fluctuations and correlation effects has an important role in determining the nature of materials.  Additional complexity arises even for single-orbital systems in connection with the spin degrees of  freedom since the electron localization itself leaves the spin entropy finite. Slater considered\cite{Slater} the possibility of antiferromagnetic (AF) symmetry breaking to release this entropy, while the AF order itself may become the origin of the insulating gap formation. Then the metal-insulator and AF transitions could be coupled each other.  Although the Mott transition is one of the  typical quantum phase transitions caused by this interplay and many studies have been done\cite{MIT}, the physical properties near the Mott transition have not been sufficiently understood in the presence of strong fluctuations.   

For the Mott transitions, there are three basic control parameters, band filling, bandwidth and dimensionality~\cite{MIT}. Although the quantum Monte Carlo (QMC) results in the two-dimensional Hubbard model have contributed in clarifying the nature of the filling-control metal-insulator transition~\cite{QMC1,QMC2,Assaad}, the method generically suffers from the negative sign problem, if the longer-ranged transfer in the model becomes large.  Recently we have developed an algorithm called path-integral renormalization group (PIRG) method\cite{PIRG,EXTRAPO}, which can be applied to any lattice structure without the sign problem. It has opened the possibility of studying correlated electron models in a much wider class. In this paper we apply PIRG method for studying the Mott transition which has not been able to be studied in other numerical methods. In particular, we study metal-insulator and magnetic transitions, in two-dimensional systems, caused by the control of the bandwidth relative to the strength of the on-site electron-electron interaction $U$.  The simplest Hubbard model on the square lattice with the nearest-neighbor transfer $t$ satisfies the perfect nesting condition and is expected to be the antiferromagnetic insulator at any nonzero $U/t$.  In real materials, however, the longer-ranged transfer is generically present in tight-binding representations, which drives a metal-insulator transition at nonzero $U/t$.  To clarify the nature of the bandwidth-control transitions, we study the phase diagram in the parameter space of the bandwidth relative to $U$ and the strength of the next-nearest neighbor transfer $t'/t$.  

Systems with the two-dimensional anisotropy are seen in many materials such as high-$T_{c}$ cuprate superconductors, organic compounds, some materials under strong electronic fields and so on.  In more general, the bandwidth-control transition is also seen
in many other systems such as V$_2$O$_3$ and RNiO$_3$~\cite{MIT}.
In this study, however, we do not focus our interest on any concrete materials and rather employ  two-dimensional single-band Hubbard models as our target to study the general physics of quantum phase transitions in electron systems. Although the Hubbard model is simple, it captures main features of the electron correlations in narrow energy bands\cite{KANA,oldHub}.  

In
$\S$\ref{SecModel}, we introduce the model for studying the Mott transition and see the physics of the non-interacting case. In $\S$\ref{SecPre}, the previous studies on this model are reviewed. In $\S$\ref{SecPIRG}, 
PIRG, which is a
numerical algorithm we use in this study, is summarized. In $\S$\ref{SecPIRGres}, we show PIRG results and determine the critical values of electron-electron interaction and discuss the nature of metal-insulator and magnetic transitions.  An important conclusion from the numerical results is the existence of a nonmagnetic insulator phase sandwiched by a paramagnetic metal and an antiferromagnetic insulator.  The Mott transition occurs at smaller $U/t$ than that of the antiferromagnetic
transition.  
We also discuss the physical properties near the two transitions.  In particular, anomalous metallic state with strong renormalization of 
carriers is suggested near the metal-insulator transition at large $U$
and large $t'/t$.
The first-order transitions are suggested both for the antiferromagnetic and metal-insulator transitions at $t'/t=0.2$. 
They are in sharp contrast with the filling-control transition where a concurrent single continuous transition at the vanishing doping concentration is observed with critical enhancement of charge and magnetic fluctuations with scaling properties.\cite{fillcon}  In $\S$\ref{conc}, we summarize our conclusions.       
\section{Model} 
\label{SecModel}  
\subsection{Hamiltonian} 
\begin{figure}  \epsfxsize=80mm  \epsffile{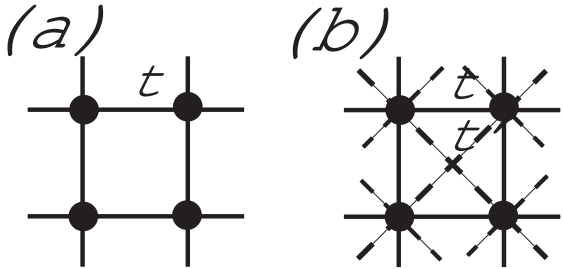}  \caption{(a)The lattice structure of the two-dimensional Hubbard  model on a square lattice. (b)The lattice structure of the two-dimensional Hubbard  model with next nearest neighbor transfers $t'$. } \label{bond} \end{figure}

In this chapter, we introduce an extended Hubbard model with longer-ranged transfers
for the study of the magnetic and metal-insulator transitions.  We use the following Hamiltonian.  
\begin{eqnarray}  \hat{H}&=&\hat{H}_{t}+\hat{H}_{U}\nonumber\\  
\hat{H}_{t}&=&-\sum_{<i,j>,\sigma}t_{ij}   
\left(c_{i\sigma}^{\dagger}c_{j\sigma}+h.c.\right)\nonumber\\  
\hat{H}_{U}&=&U\sum_{i=1}^{N}   \left(n_{i\uparrow}-\frac{1}{2}\right)   \left(n_{i\downarrow}-\frac{1}{2}\right)\nonumber\\  
t_{ij}&=&   \left\{    \begin{array}{lll}     t=1.0 \textrm{ for $(i,j)$ are the nearest neighbor sites}\\     
-t'\textrm{ for $(i,j)$ are the next-nearest neighbor sites}\\     
0\textrm{ otherwise}    
\end{array}     \right. 
\label{Hamiltonian}  
\end{eqnarray} where $i$ and $j$ represent the lattice points, $c_{i\sigma}^{\dagger}\left(c_{i\sigma}\right)$ the creation (annihilation) operator of an electron with spin $\sigma$ on the $i$-th site, $n_{i\sigma}=c_{i\sigma}^{\dagger}c_{i\sigma}$, $t_{ij}$ the transfer integral between the $i$-th site and the $j$-th site, $U$  the on-site Coulomb interaction and $N$ the number of lattice sites.  We take $t=1.0$ as the energy scale.  We study two-dimensional  square lattice with the next-nearest-neighbor transfer $t'$ illustrated in Fig.\ref{bond}.  

There are many simplifications in the Hubbard model. The most drastic one is to consider only electrons in a single orbit. However, low-energy and low-temperature properties of a simple band near the Fermi level even in the presence of a complicated structure of high-energy bands far from the Fermi level are often well described by the Hubbard model. Therefore,  the above Hubbard model has some generality in discussing  fundamental features of quantum phase transitions, which occurs at  zero temperature in the strict sense.    

\subsection{Fermi surface} 
\begin{figure}  
\hspace{-2mm}  
\begin{minipage}{.43\linewidth}   \epsfxsize=40mm   \epsffile{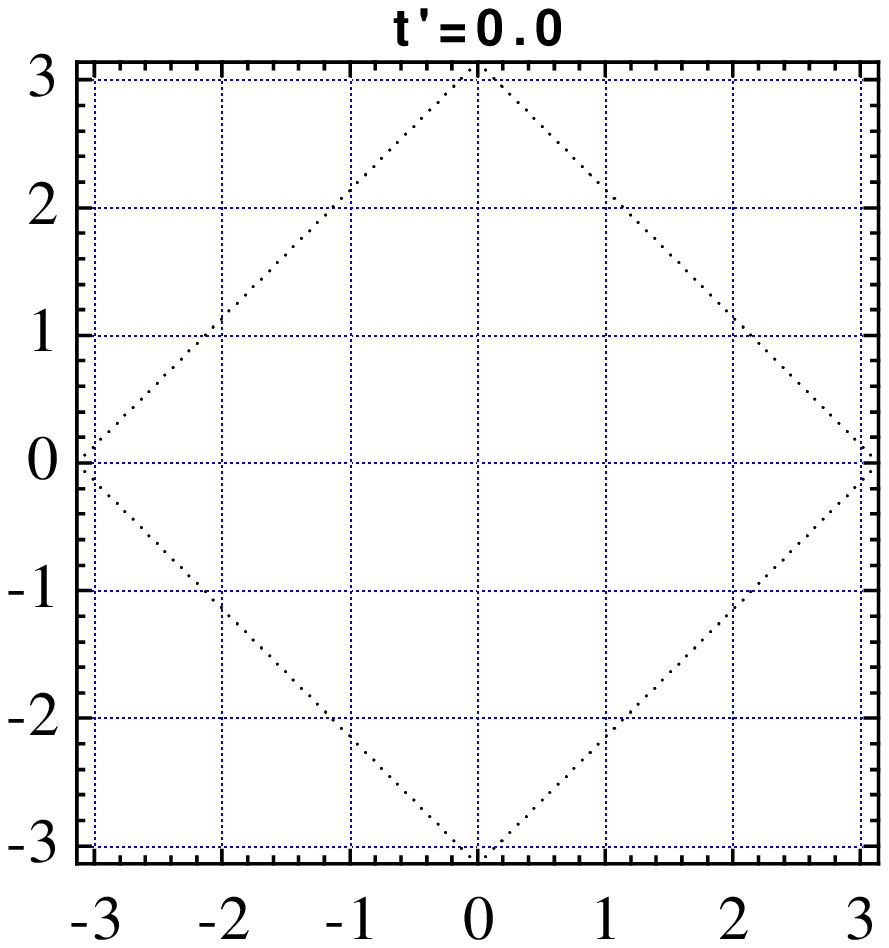}  \end{minipage}  \hspace{3mm}  
\begin{minipage}{.43\linewidth}   \epsfxsize=40mm   \epsffile{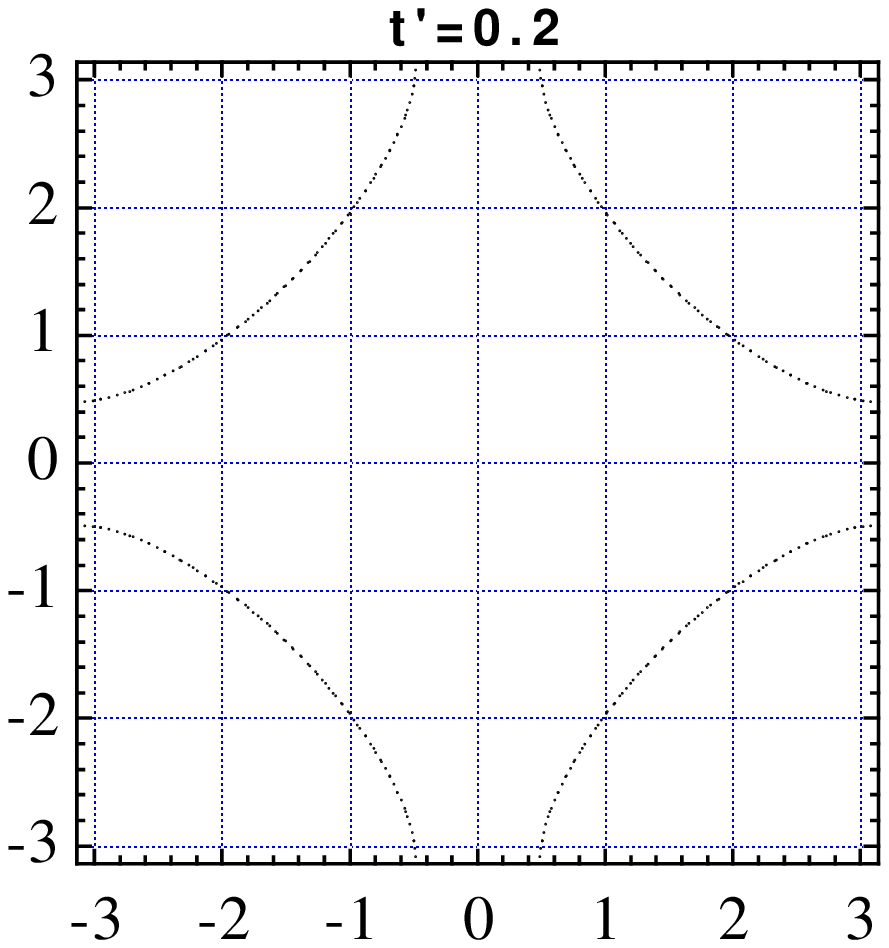}  \end{minipage}\\  \hspace{-2mm}  
\begin{minipage}{.43\linewidth}   \epsfxsize=40mm   \epsffile{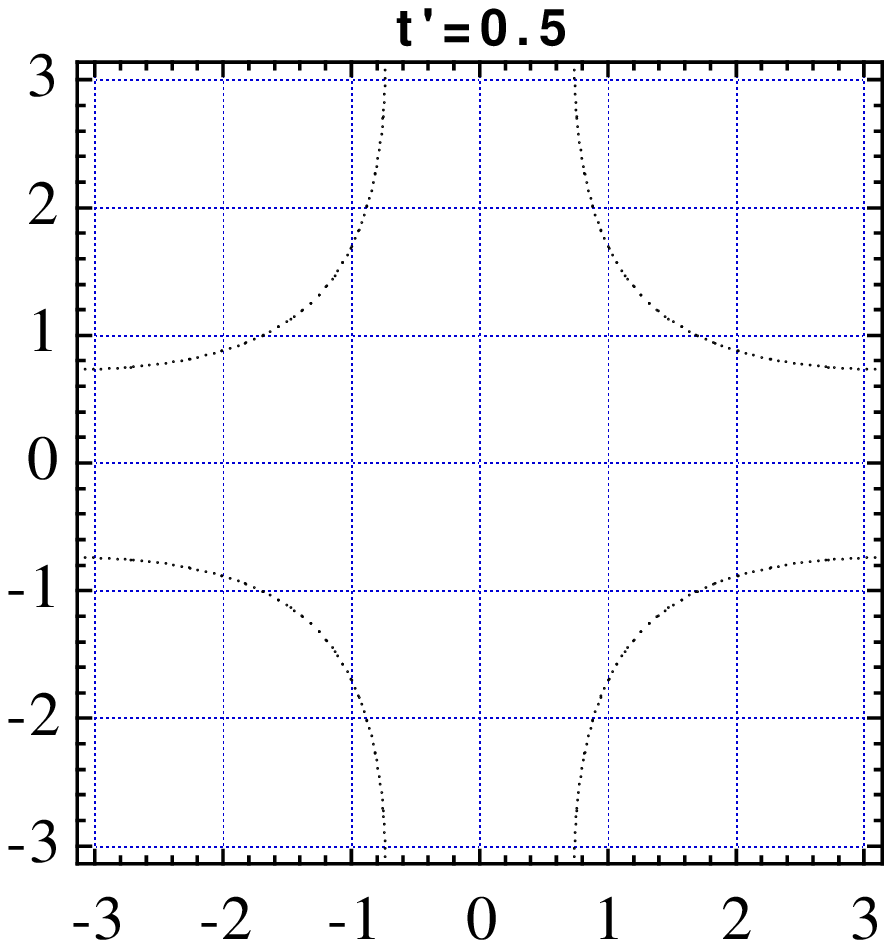}  \end{minipage}  \hspace{3mm}  
\begin{minipage}{.43\linewidth}   \epsfxsize=40mm   \epsffile{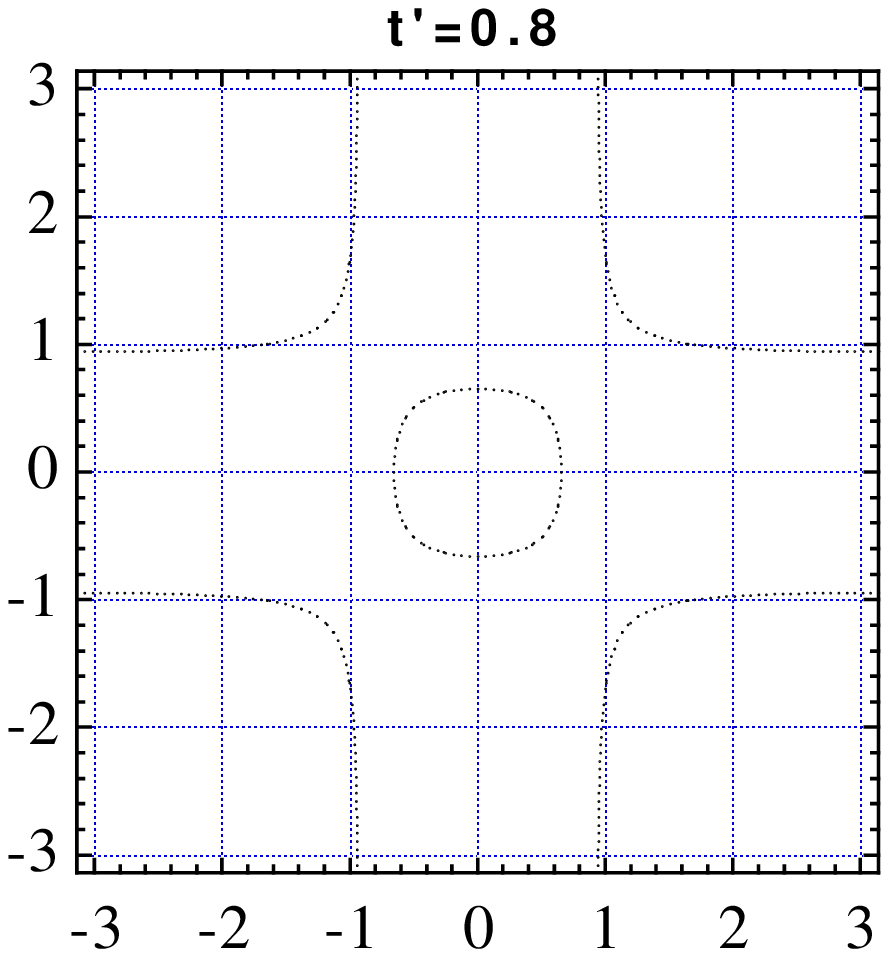}  \end{minipage}  \caption{The Fermi surface of the non-interacting Hamiltonian  $\hat{H}_{t}$}  \label{FS} \end{figure}

Before we deal with the interacting Hamiltonian $\hat{H}=\hat{H}_{t}+\hat{H}_{U}$, we study the non-interacting one $\hat{H}=\hat{H}_{t}$. The energy band is given by  \begin{equation}  E_{k}=-2t\left(\cos k_{x}a+\cos k_{y}a\right) + 4t'\cos k_{x}a \cos k_{y}a   \label{noInteraction} \end{equation} where $a$ is the lattice constant. Hereafter we take the length unit $a=1$.  
In this system the Fermi surfaces at four choices of $t'$ are shown in Fig.\ref{FS}. These are calculated on the $200\times 200$ meshed momentum plane. These figures show that the Fermi surfaces bend and the nesting is destroyed when $t'$ increases. Note that in the range where $t'$ is larger than about $0.6$, another Fermi surface appears near $(0,0)$ as shown in the $t'=0.8$ case.   

\subsection{Range of $t'$} 
\begin{figure}  
\hspace{2cm}  
\epsfxsize=80mm  
\epsffile{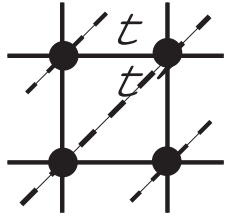}  
\caption{The lattice structure of $\kappa -(\textrm{ET})_{2}X$,  2D-organic superconductor. } 
\label{org_bond} 
\end{figure}

Some studies\cite{tprime1,tprime2,tprime} have been done to decide the three-band $d$-$p$ model  parameter by comparing a cluster calculation of the $d$-$p$ model consisting of Cu-$d_{x^{2}-y^{2}}$ and O-$p_\sigma$ orbitals and the experimental values on the high-Tc cuprate superconductors. Some studies\cite{dp_H1,Andersen} have also  been done to decide the parameter $t'$ in Hamiltonian (\ref{Hamiltonian}) by comparing the exact diagonalization results of the Hubbard model Hamiltonian (\ref{Hamiltonian}) with those of the $d$-$p$ model or by comparing tight-binding dispersion with the first-principles results. The result for $t'/t$ is on Table \ref{tprime}. $\kappa -(\textrm{ET})_{2}X$ is an organic superconductor with two-dimensional anisotropy,  whose lattice structure is different from Fig.\ref{bond} and is frequently mimicked by the structure illustrated in Fig.\ref{org_bond}. In this paper we consider the $t'=0.2$ and $t'=0.5$ cases in the lattice structure shown in Fig.\ref{bond}(b). We refer to these examples to show that our choices of $t'$ are within a realistic and sensible range as a model of several materials.  However because our interest is on metal-insulator and magnetic transitions on a general ground, we do not focus on detailed comparison with experimental results.  
\begin{table}  
\caption{The value of next nearest transfer $t'$ estimated for several  materials}   \begin{center}  
\begin{tabular}{ll}   $t'$          & material\\   $0.2$         & LSCO\\   $0.3\sim 0.5$ & YBCO,BSCCO\\   $-0.8$    & $\kappa -(\textrm{ET})_{2}X$   [ET=BEDT-TTF,X=Cu{N$(\textrm{CN})_{2}X^{\prime}$,$X^{\prime}$=Cl,Br,$\cdots$}]   
\end{tabular}  
\label{tprime}   
\end{center} 
\end{table} 
  
\subsection{Quantum phase transitions} 
In strongly correlated electron systems, metal-insulator transitions from the Mott insulator occur through three different routes\cite{MIT}.  In the extended Hubbard model described by Hamiltonian (\ref{Hamiltonian}) they are simulated in the following:  
\begin{itemize}  \item  Bandwidth control. 	
\begin{itemize} 	 \item The bandwidth is controlled by the relative 	       strength of electron-electron interaction $U/t$ and/or the 	       relative transfer $t'/t$ in the Hamiltonian (\ref{Hamiltonian}). 	\end{itemize}  \item  Band filling control. 	
\begin{itemize} 	 \item In the single band Hubbard model without the uniform 	       magnetization, the band filling  	       $n=\frac{\displaystyle{M}}{\displaystyle{N}}$ may be 	       controlled where $M$ 	       is the total number of particles and $N$ is the 	       number of the lattice sites. 	\end{itemize}  \item	Dimensionality and lattice structure control.  	
\begin{itemize} 	 \item In our model the information of the dimensionality is 	       stored in $t_{ij}$. Even if the dimension is fixed, the 	       bond structure $t_{ij}$ has a large influence on quantum 	       phase transitions.  	\end{itemize} \end{itemize} 
The corresponding controls of the above three key parameters can experimentally be achieved and designed, for example, by doping, pressure, chemical composition and magnetic fields.   Although the filling-control transition has been studied by the QMC quite extensively\cite{QMC1,QMC2,MIT,Assaad}, the nature of the bandwidth-control transition has not been studied on the same detailed level as we discussed briefly in $\S$\ref{INTRO}.   In case of the perfect nesting at $t'=0$ as shown in Fig.\ref{FS}, the critical value $U_{c}$ for the bandwidth-control transition 
is believed to be zero.  It is also believed that the critical value $U_{c}$ is nonzero at $t'\not= 0$ because of the destruction of the nesting. 
Then the bandwidth-control transition can only be studied at $t'\not= 0$. It is difficult, however, to apply QMC in $t'\not= 0$ regime because of the sign problem.   

Since the bandwidth control is one of the main routes of the metal-insulator transition in many materials as discussed in 
\S 1, it is desired to clarify its general feature. Here we focus our study on the bandwidth-control transition since PIRG for the first time makes it  possible to study it numerically.    More concretely, we fix the band filling at $n=1$ and study two cases of the relative transfers $t'/t=0.2$ and $0.5$ as we discussed. We search the quantum phase transition point by changing the relative  strength of correlations $U/t$. We determine the critical values $U_{c}$ of the transitions scaled by $t=1.0$.  We study two quantum phase transitions, one the metal-insulator transition and the other, magnetic transition. The former is studied by estimating the $U$ dependence of the double occupations and the charge excitation gap while the latter by measuring the equal-time spin-spin correlations.          
\section{Previous studies on this model} 
\label{SecPre} 
There exist some studies on the Hubbard model (\ref{Hamiltonian}) in the literature.  The critical values $U_{c}$ determined by these previous studies are on Table \ref{UC02} and Table \ref{UC05}.  The Hartree-Fock approximation is the simplest way to approach this issue\cite{HF}.  Kondo and Moriya have used the Hartree-Fock approximation for the two-dimensional model and calculated the free energy as a function of the uniform  staggered moment $m=|\langle n_{j\uparrow}\rangle -\langle n_{j\downarrow}\rangle|/2$ and the strength of interaction $U$.  They have searched the case of $U$ in which $m\ne 0$ gives the lowest energy,  to discuss an antiferromagnetic transition with the periodicity given by the wavenumber $(\pi,\pi)$. To discuss metal-insulator transitions, they have studied the presence of the charge excitation gap in the density of states.  They have also studied the same system by using the Green's function approach with the self-energy up to the second order in the strength of the interaction $U$ \cite{HF2}.    

There exist also several studies on the paramagnetic state and the antiferromagnetic state by the Gutzwiller approximation.\cite{GA1,GA2,VMC1,VMC2}  It was claimed\cite{HF} that since difference in energy is small between the antiferromagnetic states estimated by the Hartree-Fock approximation and that by the Gutzwiller approximation, it would be possible to estimate the critical value of the magnetic transition by using the paramagnetic state of Gutzwiller approximation and the antiferromagnetic state of Hartree-Fock approximation.  
A general tendency of the Hartree-Fock 
approximation shows the existence of the antiferromagnetic metal phase. 
However, the Hartree-Fock and Gutzwiller approximations may overestimate the 
region of symmetry-broken states such as the antiferromagnetically ordered
phase.  Therefore, careful studies are desired to determine the phase 
diagram by considering quantum fluctuation effects. 

 The method to discuss Mott transitions by QMC shares similarity to the method we use in PIRG study shown in the next section.  Quantum Monte Carlo suffers from the sign problem in this regime of $t'/t$.  As far as we know, no QMC studies for the bandwidth-control metal-insulator transition are available.  There are some QMC studies for magnetic transitions  in weakly frustrated regime.\cite{QMC2,Hirsch}  The magnetic transition has been discussed by the equal-time spin correlations only for the possibility of the antiferromagnetic order at the wave number $(\pi,\pi)$ while the possibility for other magnetic states has not been examined.  The definition of equal-time spin correlation is shown in the next section.   
\begin{table}[htbp]  
\caption{Critical value $U_{c}/t$ in case of $t'/t=0.2$}  
\begin{tabular}{@{\hspace{\tabcolsep}\extracolsep{\fill}}ccc}   \hline    & magnetic transition & Metal-Insulator transition \\   \hline   Hartree Fock \cite{HF} & $2.064$ & $2.064$ \\   Green's function \cite{HF2} & $4.00$ & $4.00$ \\   Gutzwiller \cite{HF} & $3.920$ & no data \\   Quantum Monte Carlo \cite{Hirsch} & $2.5\pm 0.25$ & no data \\   \hline  
\end{tabular}  
\label{UC02} 
\end{table} 
\begin{table}[htbp]  
\caption{Critical value $U_{c}/t$ in case of $t'/t=0.5$}  
\begin{tabular}{@{\hspace{\tabcolsep}\extracolsep{\fill}}ccc}   \hline    & magnetic transition & Metal-Insulator transition \\   \hline   Hartree Fock \cite{HF} & $3.215$ & $3.290$ \\   Gutzwiller \cite{HF} & $5.259$ & no data \\   \hline  
\end{tabular}  
\label{UC05} 
\end{table}     
\section{Path-integral renormalization group} 
\label{SecPIRG} 
Path-integral renormalization group (PIRG) is a numerical algorithm for studying the ground state properties.\cite{PIRG}  The process projecting onto the ground state $|\psi_{g}\rangle$  is performed in the imaginary time direction by the Hamiltonian operator 
$\hat{H}$ as, \begin{displaymath}  |\psi_{g}\rangle=\lim_{\tau \to \infty} \exp[-\tau \hat{H}]|\phi_{initial}\rangle . \end{displaymath} Therefore its formalism can be applied to any kind of systems and there is no restriction on the spatial dimension of the system. In this path-integral formalism, the ground state is represented by a set of chosen basis states $|\phi_i\rangle$, \begin{displaymath}  |\psi_{g}\rangle=\sum_{i}c_{i}|\phi_{i}\rangle . \end{displaymath} By the numerical renormalization, relevant basis states are selected and irrelevant basis states are integrated out. This makes it possible to calculate an approximate ground state  within a fixed dimension, $L$, of the Hilbert space,  directly as an optimized linear combination of the numerically 
chosen basis states:  
\begin{displaymath}  
|\psi_{g}\rangle\approx\sum_{i=1}^{L}w_{i}|\phi_{i}\rangle . 
\end{displaymath} 
In this process, the negative sign problem does not appear even in generic fermion models. In this way, PIRG can be applied to the systems which can not be treated by existing algorithms such as quantum Monte Carlo method or the density matrix renormalization group.   

Because converged states by PIRG are an approximate ground states under  the restriction on the number of the basis states, the properties in the exact ground state can  be achieved by the extrapolation of the number of states $L$ to the dimension of the whole Hilbert space. We have shown a general extrapolation procedure\cite{EXTRAPO} by using the relation:   \begin{displaymath}  \langle\hat{H}\rangle - \langle\hat{H}\rangle_{g} \propto \Delta E \end{displaymath} where $\Delta E$ is the energy variance, 
\begin{displaymath}  
\Delta E=\frac{\langle\hat{H}^{2}\rangle - \langle\hat{H}\rangle^{2}}   {\langle\hat{H}\rangle^{2}}, \end{displaymath} 
$\langle\quad\rangle_{g}$, the expectation value in the true ground state  and $\langle\quad\rangle$, the expectation value in an approximate ground state. This relation holds for sufficiently converged approximate state.  When we use the Slater determinants as the basis states $|\phi_{i}\rangle$, we confirm that the results with $L$ roughly more than a hundred follow the above relation and a linear extrapolation can be used in the Hubbard model with the parameter values we employ in this paper. On the physical quantity $\hat{A}$, a similar relation holds in most cases.\cite{EXTRAPO}  
\begin{displaymath}  
\langle\hat{A}\rangle - \langle\hat{A}\rangle_{g} \propto \Delta E . 
\end{displaymath} 
For short-ranged correlation functions, the linearity holds at the same level as the energy in the thermodynamic limit~\cite{EXTRAPO}.  By these extrapolation procedures, more accurate results are obtained.  We note that, although the results at finite $L$ satisfy the variational principle, it is not strictly satisfied for the extrapolated results.   

Error bars of estimated physical quantities in our method have basically 
two origins.  One comes from the extrapolation procedure from finite $L$ 
to the dimension of the whole Hilbert space.  If the linearity in the 
fitting as a function of the energy variance becomes worse, the error bars 
increase.  The other comes from the extrapolation to the thermodynamic limit
by assuming finite size corrections.  The latter is a common origin of error bars 
to all the other numerical methods for finite size systems while the former 
is specific to this PIRG method.  The error bars in our analyses are given 
from the combination of these two types of error bars.          
\section{PIRG results}
\label{SecPIRGres}
\subsection{Method for analyzing transitions}

\subsubsection{$U$ dependence of energy}
When a transition caused by the on-site interaction $U$ is of the 
first-order, $\partial E_{g}/\partial U$ should have a jump at the transition
point $U_{c}$ because of the level crossing of the ground state as a
function of $U$. 
Here $E_{g}$ is the ground state energy, namely $\langle \hat{H}\rangle_{g}$. 
Even if the transition is of higher-order, some singularities of the
energy may be seen at $U_{c}$ in general. 
From Hamiltonian (\ref{Hamiltonian}), $\partial E_{g}/\partial U$ can be
transformed as
\begin{equation}
 \frac{\partial E_{g}}{\partial U} 
  = \left\langle \sum_{i=1}^{N}n_{i\uparrow}n_{i\downarrow}
     - \left(M-\frac{N}{4}\right) \right\rangle_{g},
\end{equation}
where $M=M_{\uparrow}+M_{\downarrow}$ is the particle number.
Then, we can determine the transition point by estimating the
double occupancy:
\begin{equation}
 \langle n_{\uparrow}n_{\downarrow} \rangle_{g}
  \equiv \frac{1}{N}\sum_{i=1}^{N}\left\langle n_{i\uparrow}n_{i\downarrow}
				   \right\rangle_{g}.
\end{equation}
In our calculation, the double occupancy is obtained after the extrapolation 
to the thermodynamic limit by the fitting in the form 
$ \langle n_{\uparrow}n_{\downarrow} \rangle_{g}(N=\infty)=
\langle n_{\uparrow}n_{\downarrow} \rangle_{g}(N=N) + \alpha/\sqrt{N}$.
The data we have obtained for the double occupancy indeed well follow 
this finite-size correction as in energy. 

\subsubsection{Metal-insulator transitions}

Metal-insulator transitions can be discussed by computing the presence of a jump
of the chemical potential at half filling. Because the chemical
potential is defined by $\mu\left(n\right)=\partial E/\partial n$ where
$n=\left(M_{\uparrow}+M_{\downarrow}\right)/N$ is the particle density,
we calculate the chemical potential by the following equation:
\begin{equation}
 \mu\left(\frac{2M-1}{N}\right) 
 = \frac{\Delta E}{\Delta n}
 =\frac{E_{g}\left(M,M\right)-E_{g}\left(M-1,M-1\right)}{2} 
\end{equation}
where $E_g\left(M_{\uparrow},M_{\downarrow}\right)$ is the ground state
energy of the whole system including $M_{\uparrow}$ electrons with the
up spin and $M_{\downarrow}$ electrons with the down spin. The charge
excitation gap, if it exists at half filling, is calculated as
\begin{equation}   
 \Delta_{c}=\frac{1}{2}\left[\mu\left(\frac{N+1}{N}\right)
			-\mu\left(\frac{N-1}{N}\right)\right] .
			\label{Gscale}
\end{equation}
Namely, we calculate the ground state energy $E_{g}\left(N/2-1,N/2-1\right)$,
$E_{g}\left(N/2,N/2\right)$, $E_{g}\left(N/2+1,N/2+1\right)$ by PIRG
after extrapolation to the thermodynamic limit. 
We estimate the charge excitation gap from Eq.(\ref{Gscale}). 
We note here that the accuracy becomes worse for the estimate of the
charge-excitation gap because we have to perform the second-order
differentiation numerically from the ground state energies 
for the particle numbers $2(M+1),2M$ and $2(M-1)$. 
We use a finite-size scaling function in the form  
\begin{equation}
 \Delta_{c}\left(N\right)=\Delta_{c}+\frac{1}{\sqrt{N}}\Delta^{\prime}
  \label{finite}
\end{equation}
where $N$ is the number of sites in two-dimensional systems. 
We take the above scaling form because the finite-size effect for the
Hartree-Fock SDW gap equation is given by the series of $1/\sqrt{N}$. 
Equation (\ref{finite}) is the same as that used in the
literature\cite{QMC2}. 

In addition to the charge excitation gap, 
we also calculate the momentum distribution to discuss the electronic
structure of systems. It is defined by 
\begin{equation}
 n \left(\mib{q}\right)=\frac{1}{2N}\sum_{i,j}^{N}\left\langle
    c_{j\uparrow}^{\dagger}c_{i\uparrow}+c_{j\downarrow}^{\dagger}c_{i\downarrow}
    \right\rangle e^{i\mib{q}\left(\mib{R}_{i}-\mib{R}_{j}\right)}
    \label{momdisEq}
\end{equation}
where $\mib{R}_{i}$ is the vector representing the place of the $i$-th
site. 

\subsubsection{Magnetic transitions}
Magnetic transitions can be discussed by the
equal-time spin correlations in the momentum space, 
$S(\mib{q})$, defined by  
\begin{equation}
 S\left(\mib{q}\right)=\frac{1}{3N}\sum_{i,j}^{N} 
  \left\langle\mib{S}_{i}\mib{S}_{j}
  \right\rangle e^{i\mib{q}\left(\mib{R}_{i}-\mib{R}_{j}\right)}
  \label{spinco}
\end{equation}
where $\mib{S}_{i}$ is the spin of the $i$-th site and $\mib{R}_{i}$ is
the same as that used in Eq.(\ref{momdisEq}). 
Each element of the spin is defined by 
\begin{eqnarray}
 S_{i}^{x}&=&\frac{1}{2}\left(S_{i}^{+}+S_{i}^{-}\right)
          =\frac{1}{2}\left(c_{i\uparrow}^{\dagger}c_{i\downarrow}
		             + c_{i\downarrow}^{\dagger}c_{i\uparrow}\right)
	  \nonumber\\
 S_{i}^{y}&=&\frac{1}{2i}\left(S_{i}^{+}-S_{i}^{-}\right)
          =\frac{1}{2i}\left(c_{i\uparrow}^{\dagger}c_{i\downarrow}
		             - c_{i\downarrow}^{\dagger}c_{i\uparrow}\right)
	  \nonumber\\
 S_{i}^{z}&=&\frac{1}{2}\left(n_{i\uparrow}-n_{i\downarrow}\right) .
  \label{detailspinco}
\end{eqnarray}
If a long-range magnetic order exists 
with the momentum $\mib{q}_{peak}$ in the
thermodynamic limit, it can be extracted from the extrapolation of
$S(\mib{q}_{peak})$ in finite-size systems. 
From the theory of spin wave in two dimensions, the following scaling is expected:
\begin{equation}
 S(\mib{q}_{peak})=Nm^{2}+S_{c}\sqrt{N}
  \label{FSS}
\end{equation}
where $m$ is the staggered magnetization 
\begin{equation}
 m = \frac{1}{N}\sum_{i}\exp (i\mib{q}_{peak}\mib{r}_i)
  \langle n_{i\uparrow}-n_{i\downarrow}\rangle ,
\end{equation}
$N$, the number of lattice sites of the two-dimensional system, $S_{c}$, 
the short-ranged part of the structure factor and $(x_{i},y_{i})$, the
vector indicating the coordinate of the $i$-th site. 
Then if the long-range order exists at $\mib{q}_{peak}$, 
we expect $S(\mib{q}_{peak})/N$ to follow a linear function
of $1/\sqrt{N}$ with a nonzero offset at $1/\sqrt{N}=0$. 
 
\subsection{$t'/t=0.2$ case}

\subsubsection{Metal-insulator transition at $t'/t=0.2$}

\begin{figure}
 \epsfxsize=80mm
 \epsffile{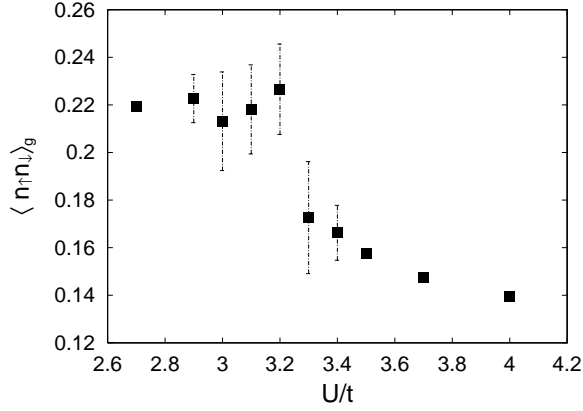}
 \caption{The expectation value of the double occupancy 
$\langle n_{\uparrow}n_{\downarrow}\rangle_{g}$ after finite-size scaling. 
The plots are for $t'/t=0.2$.}
\label{DOB02}
\end{figure}

Figure \ref{DOB02} shows the double occupancy as a function of the
relative interaction $U/t$ after the size extrapolation to the
thermodynamic limit. It suggests a jump of $\langle
n_{\uparrow}n_{\downarrow}\rangle$ between $U/t=3.2$ and $3.3$. 
Although the possibility of a very sharp second-order transition can not
be excluded, a first-order transition seems to occur at
$U/t=U_{c1}/t=3.25\pm 0.05$. The jump of the double occupation 
$\langle n_{\uparrow}n_{\downarrow}\rangle$ is estimated to be 
roughly between 0.04 and 0.05. 

Here, we discuss the origin of relatively large error bars near the 
transition points in Fig.~\ref{DOB02}.  Near the first-order
transition, fluctuations are in general not enhanced. Therefore,
one might argue that the large error bars observed in Fig. \ref{DOB02}
as compared to Fig. 12 would contradict this general expectation.  In the
present case, however, the large error bars are actually generated from the 
character of the first-order transition itself.  The level crossing
between metallic and insulating phases seem to occur at different $U_c$
for different system sizes.  $U_c$ seems to increase with increasing 
system sizes. This is natural because  the imposed periodic boundary 
condition appears to
suppress charge fluctuations in smaller system sizes. Then, if we fix $U$,
there exists a region where the insulating behavior at smaller system 
size is transformed to a metallic state at larger system sizes.  
In fact, for example, the ground state at $U=3.0$ indicating a signature of
insulating behavior with $\langle n_{\uparrow}n_{\downarrow}\rangle \sim
0.15$ at the 6 by 6 lattice transforms to a metallic behavior with 
$\langle n_{\uparrow}n_{\downarrow}\rangle \sim 0.2$ at the 10 by 10 lattice.
This transition makes the extrapolation difficult and makes the errors 
large.     

\begin{figure}
 \epsfxsize=80mm
 \epsffile{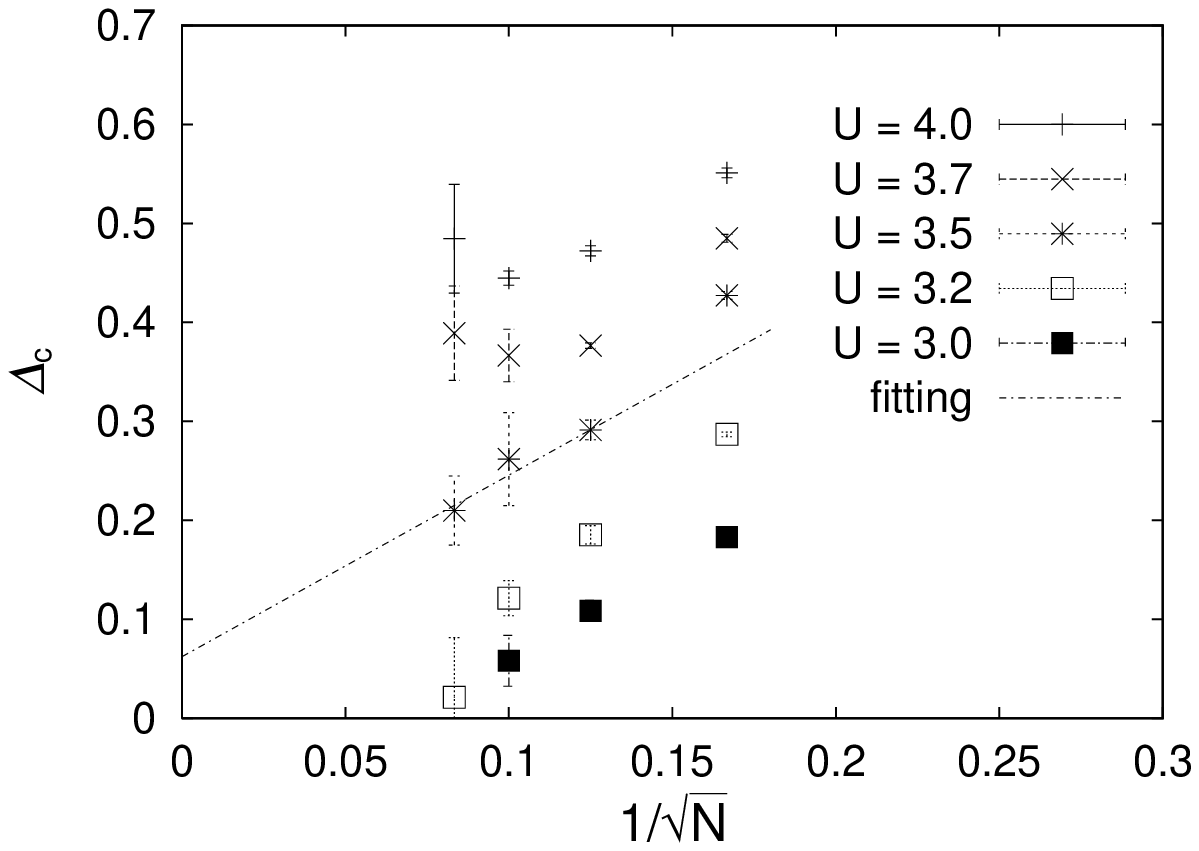}
 \caption{Finite-size scaling of $\Delta_{c}$ for some choices of the
 strength of the electron-electron interaction $U$ in $t'/t=0.2$ lattice. }
\label{Fgap02}
\end{figure}

Figure \ref{Fgap02} shows the charge excitation gap near this
transition. This figure suggests that the charge excitation gap around
 $0.1$ exists at $U/t=3.5$ and does not exist at $U$ lower than
$3.2$ in agreement with the above analysis by $\partial E_{g}/\partial U$. 
Therefore we conclude that the first-order transition shown in
Fig.\ref{DOB02} is a metal-insulator transition. 
A sharp drop of the gap with increasing system size for $U/t=3.0$ and
3.2 is related with the tranformation from the insulating to the metallic branch
mentioned above. 

Here we estimate the validity of the value of the charge excitation gap. 
From Eq.(\ref{Gscale}), the charge excitation gap $\Delta_{c}$ is equal
to the half of the sum of the excitation energies for making a holon and
a doublon in the insulating state. 
Consequently, if we add a doubly occupied site in a half-filled
system, the excitation energy is equal to $2\Delta_{c}$. 
Then the total energy cost caused by the jump of averaged double occupancy
in the system to make a metallic state from an insulating ground state 
is estimated to be roughly 
$2N\delta\langle n_{\uparrow}n_{\downarrow}\rangle\Delta_{c}$,
where $\delta\langle n_{\uparrow}n_{\downarrow}\rangle$ denotes the jump
of $\langle n_{\uparrow}n_{\downarrow}\rangle$ between the insulating and metallic states. 
Since this estimate does not include the interaction energy among
holons and doublons, this energy cost may be the lower bound for the true
energy cost to make a metal from the insulating ground state for $U>U_{c1}$. 
The true energy cost can be computed roughly as $(U-U_{c1})\times\delta\langle
n_{\uparrow}n_{\downarrow}\rangle$ which is 
proportional to  $(U-U_{c1})\times |[\partial E_{g}/\partial U]_{metal}-
[\partial E_{g}/\partial U]_{insulator}|$. Therefore, 
\begin{eqnarray}
 2N\delta\langle n_{\uparrow}n_{\downarrow}\rangle\Delta_{c}
  < N(U-U_{c1})\times\delta\langle n_{\uparrow}n_{\downarrow}\rangle\nonumber \\
 \Leftrightarrow 
  \Delta_{c}<\frac{U-U_{c1}}{2}
  \label{gapEST}
\end{eqnarray}
From this equation, the charge excitation gap $\Delta_{c}$ at $U/t=3.5$ is
estimated to be less than $(3.5-3.3)/2=0.1$.
The result shown in Fig.\ref{Fgap02} satisfies this rough estimate. 
It should be noted that even in the first-order transition, the 
charge excitation gap opens continuously from zero for $U>U_{c1}$
as we see from Eq.~\ref{gapEST}. 

In addition to that, 
the behavior of the momentum distribution at half filling shown in
Fig.\ref{momdis02} lends another support to the above
discussion. 
At $U/t=3.2$, the momentum
distribution is close to $1.0$ inside the Fermi surface of the
non-interacting system shown in Fig.\ref{FS} and nearly $0$ outside. 
At $U$ larger than $3.5$ the jump smears and the ground state of the
system appears to change to a non-Fermi-liquid state. 
Similar behaviors are shown for the simple square-lattice
Hubbard model in Fig.20 of Ref. \cite{QMC1}.  
Because we do not know whether a metallic state which is not the 
Fermi liquid exists or not and can not determine the jump of the
momentum distribution precisely, Fig.\ref{momdis02} does not
directly justify that the system is in an insulating phase for $U/t\geq
3.5$.  
However, the figure indicates a qualitative change in the electronic
structure around $U_{c1}=3.25\pm 0.05$.
%
\begin{fullfigure}  
\hspace{4cm}  
\begin{minipage}{.55\linewidth}   \epsfxsize=80mm   \epsffile{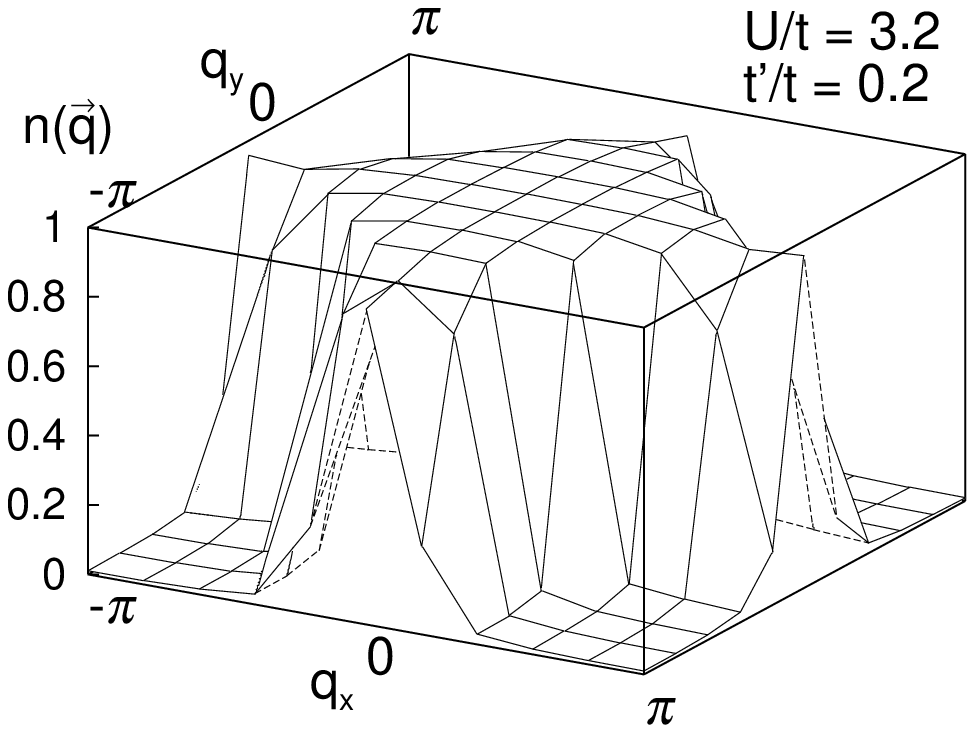}  
\end{minipage}\\  
\begin{minipage}{.55\linewidth}   \epsfxsize=80mm   \epsffile{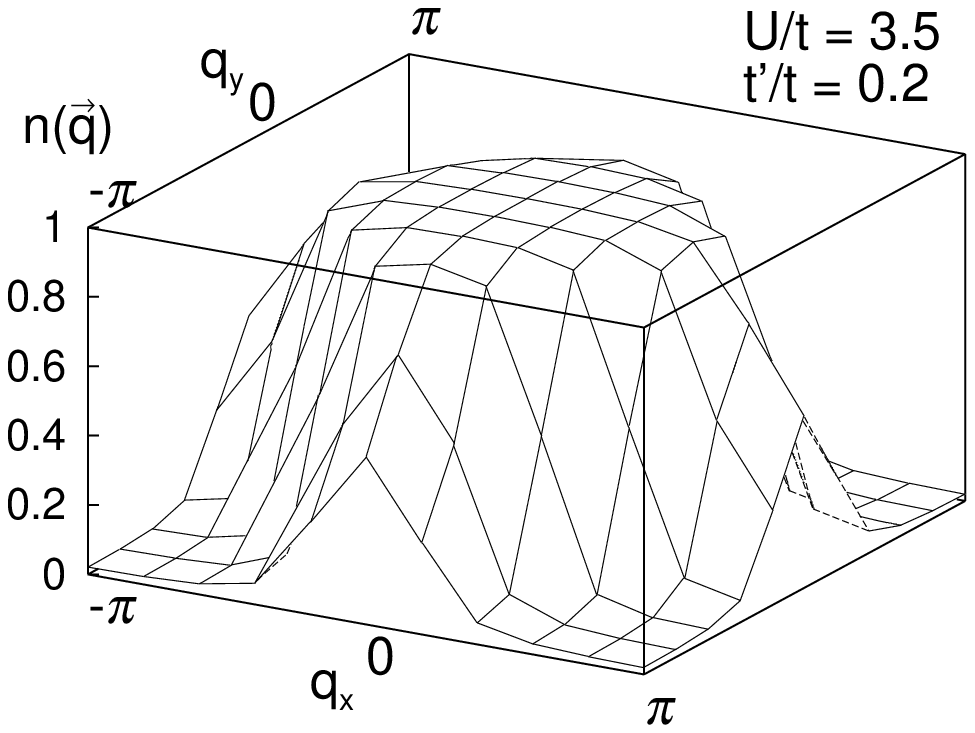}  
\end{minipage}  
\hspace{3mm}  
\begin{minipage}{.55\linewidth}   \epsfxsize=80mm   \epsffile{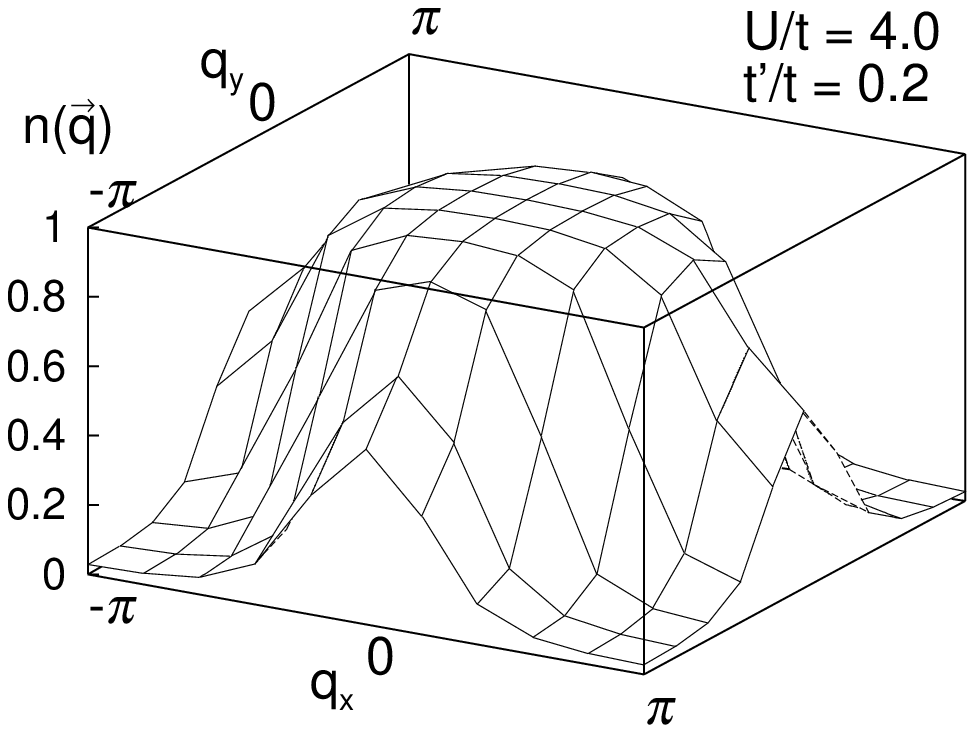}  
\end{minipage}  
\caption{The momentum distribution on $10\times 10$ lattice at  half filling for $t/t'=0.2$}  
\label{momdis02} 
\end{fullfigure} %

\subsubsection{Magnetic transition at $t'/t=0.2$} 
Examples of the structure of the equal-time spin correlations are shown in Fig.\ref{spinstructure}.  
The peak of $S(\mib{q})$ at half filling is always at $(\pi,\pi)$ in  the momentum plane.  
Therefore we study the antiferromagnetic correlation at wavelength $(\pi,\pi)$ for $t'/t=0.2$.  
 
\begin{figure}  
\epsfxsize=80mm  
\epsffile{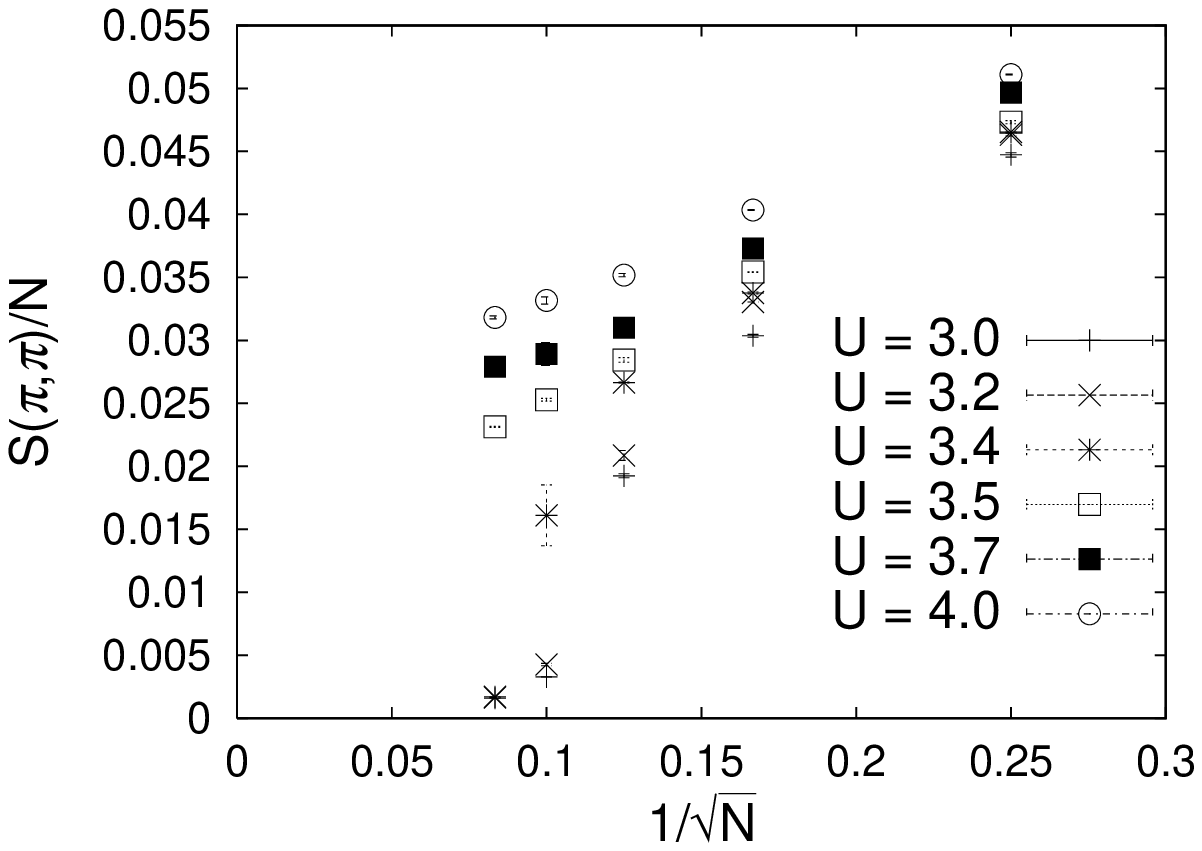}  
\caption{Finite-size scaling of $S\left(\pi,\pi\right)$ for some  choices of the strength of the electron-electron interaction $U$ in  half-filled $t'/t=0.2$ lattice systems. } 
\label{H02} 
\end{figure}

The finite-size-scaling extrapolation of $S(\pi,\pi)$ is shown in Fig.\ref{H02}.  
We also show finite-size scaling in other type of plots.  
$S(\pi,\pi)$ goes to infinity in the thermodynamic limit when the antiferromagnetic order exists.  Figures \ref{I02} and \ref{invI02} show these behaviors. From these finite-size scaling, the magnetic transition is estimated to occur at $U_{c2}/t=3.45\pm 0.05$ for $t'/t=0.2$.  

\begin{figure}  
\epsfxsize=80mm  
\epsffile{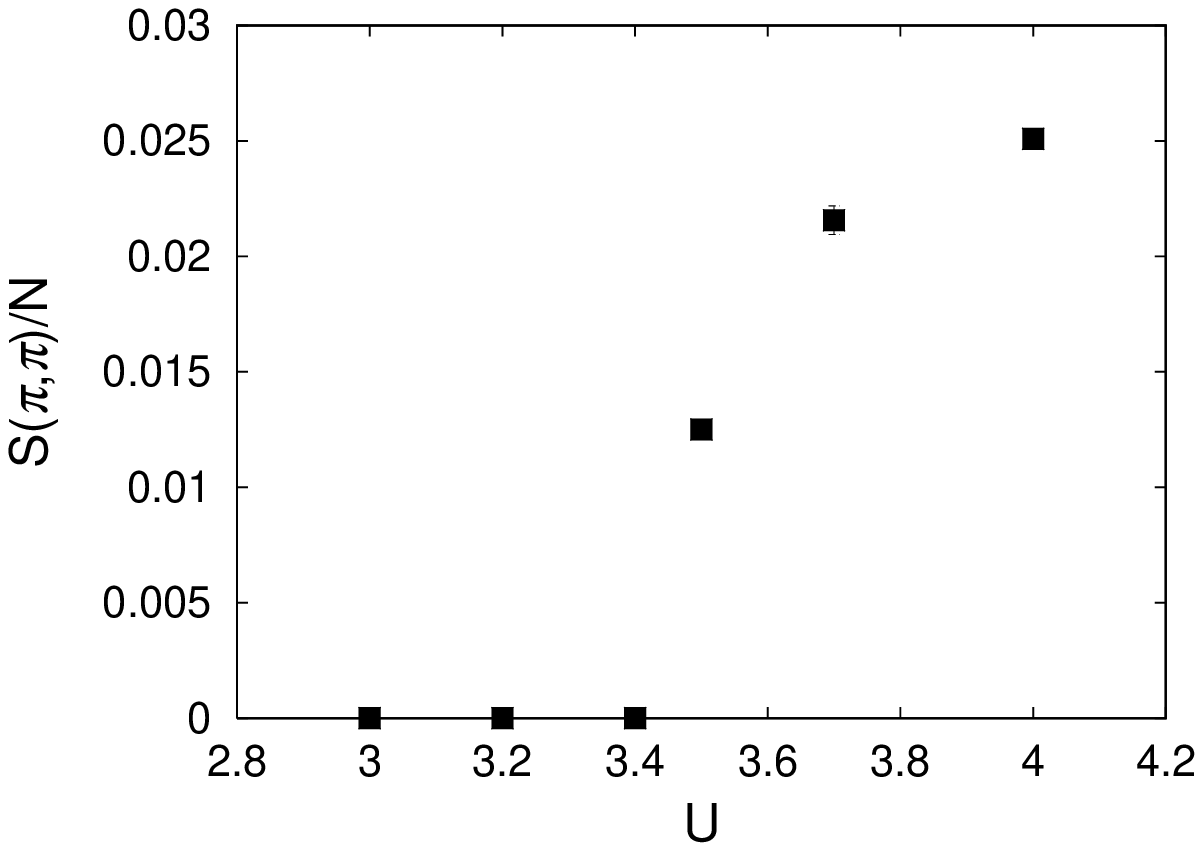}  
\caption{$S\left(\pi,\pi\right)/N$ in the thermodynamic limit and as a  function of the strength of the electron-electron interaction $U$ in  half-filled $t'/t=0.2$ lattice systems. } 
\label{U_S} 
\end{figure}

\begin{figure}  
\epsfxsize=80mm  
\epsffile{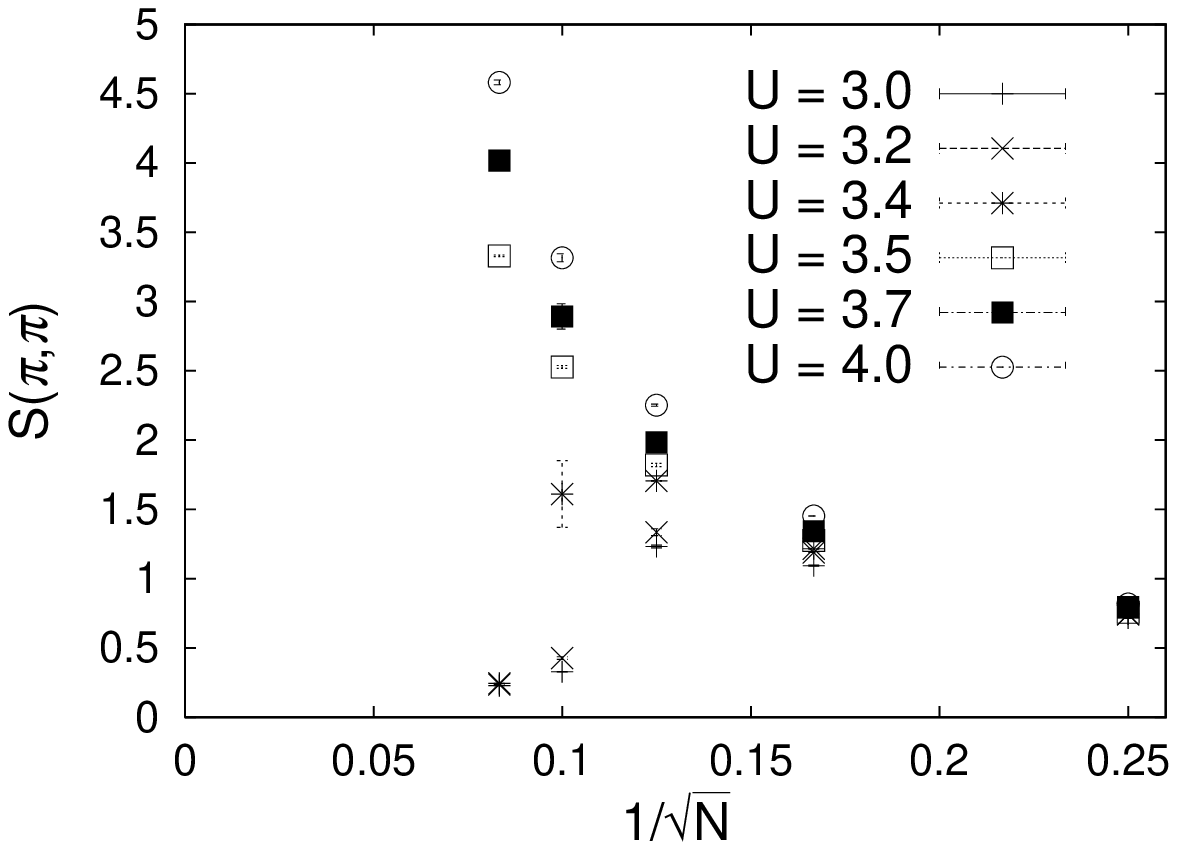}  
\caption{Finite-size scaling of $S\left(\pi,\pi\right)$ for some  choices of the strength of the electron-electron interaction $U$ in  half-filled $t'/t=0.2$ lattice systems.  } 
\label{I02} 
\end{figure}

\begin{figure}  
\epsfxsize=80mm  \epsffile{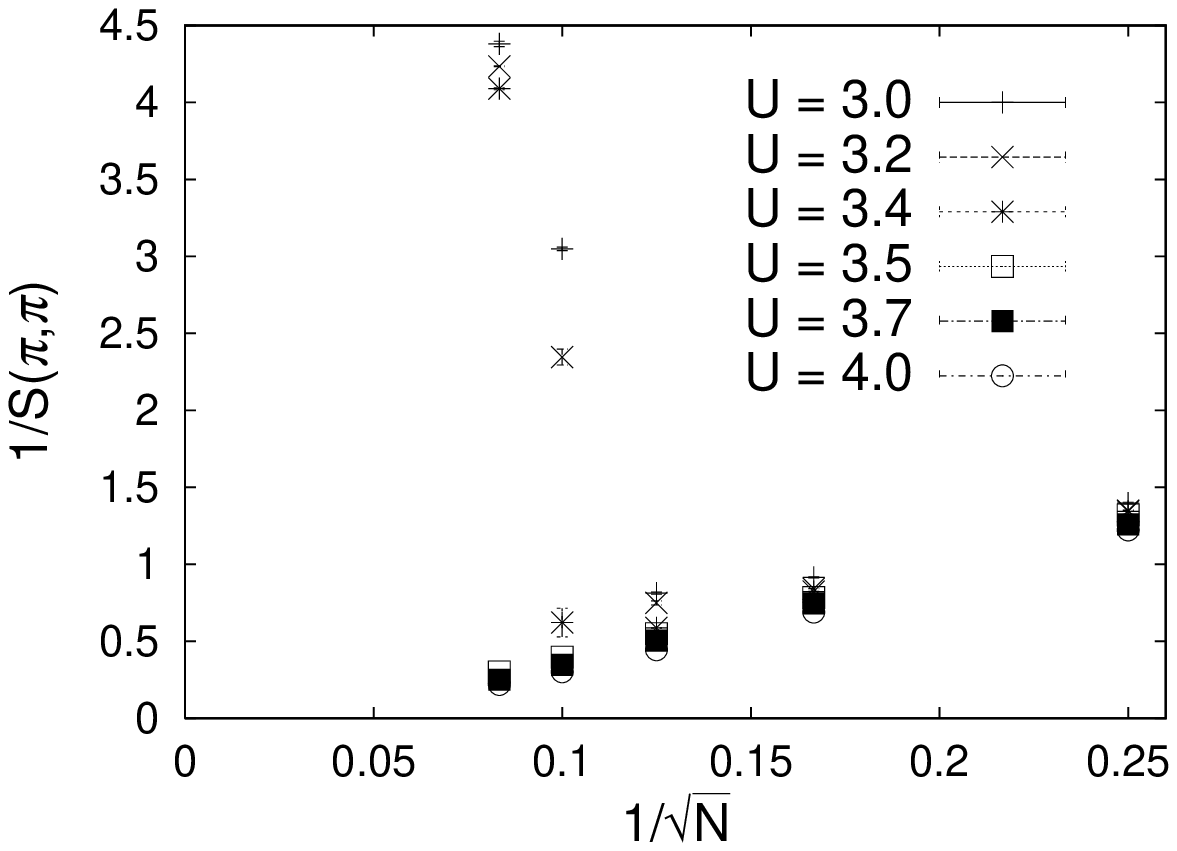}  
\caption{Finite-size scaling of $S\left(\pi,\pi\right)$ for some choices  of the strength of the electron-electron interaction $U$ in half-filled  $t'/t=0.2$ lattice systems.  } 
\label{invI02} 
\end{figure}

This value is substantially larger than  $U_{c2}/t=2.5\pm 0.25$ claimed from QMC study\cite{Hirsch}.  Because systems only up to $N=64$ have been studied by QMC in that study  and Fig.\ref{H02} shows the sharp decreases of $S(\pi,\pi)$ on the systems larger than $8\times 8$, we think that our estimate gives better accuracy and the true transition point $U_{c2}/t$ is at $3.45\pm 0.05$ for $t'/t=0.2$. 
A sharp decrease of $S(\pi,\pi)$ is again related with the 
finite size effects accompanied with the first-order magnetic transition: 
For $U/t=3.0$ and 3.2 the smaller systems show the behavior of AF ordered 
state, while it is lost in larger system sizes. 
The first-order character of the magnetic transition is supported from the following two analyses.  One is the presence of a sharp decrease of the magnetization at the  
transition. Figure~\ref{U_S} shows the results after the finite-size scaling extrapolation for $S(\pi,\pi)/N$, which is equal to $m^{2}$.  This suggests either a sharp continuous transition
or the behavior of the staggered 
magnetization $m$ at the first-order transition.  The second support is in Figs.\ref{I02} and \ref{invI02} which show that when the electron-electron interaction $U$ increases, $S(\pi,\pi)$ goes to diverge suddenly between $U/t=3.4$ and $U/t=3.5$, not continuously.  The data at $U/t=3.0$ ,$3.2$ and $3.4$ do not show growth of short-ranged correlation at the largest system size. 

  Although we do not detect a jump of $\partial E_{g}/\partial U$ at this first-order magnetic transition in Fig.\ref{DOB02}, it is likely  that the energy difference between paramagnetic insulator and antiferromagnetic insulator is far less than that between paramagnetic metal and paramagnetic insulator and then a jump of $\partial E_{g}/\partial U$ is too small to be detected in the present accuracy of 
calculations.  
\begin{fullfigure}  
\begin{minipage}{.55\linewidth}   \epsfxsize=80mm   \epsffile{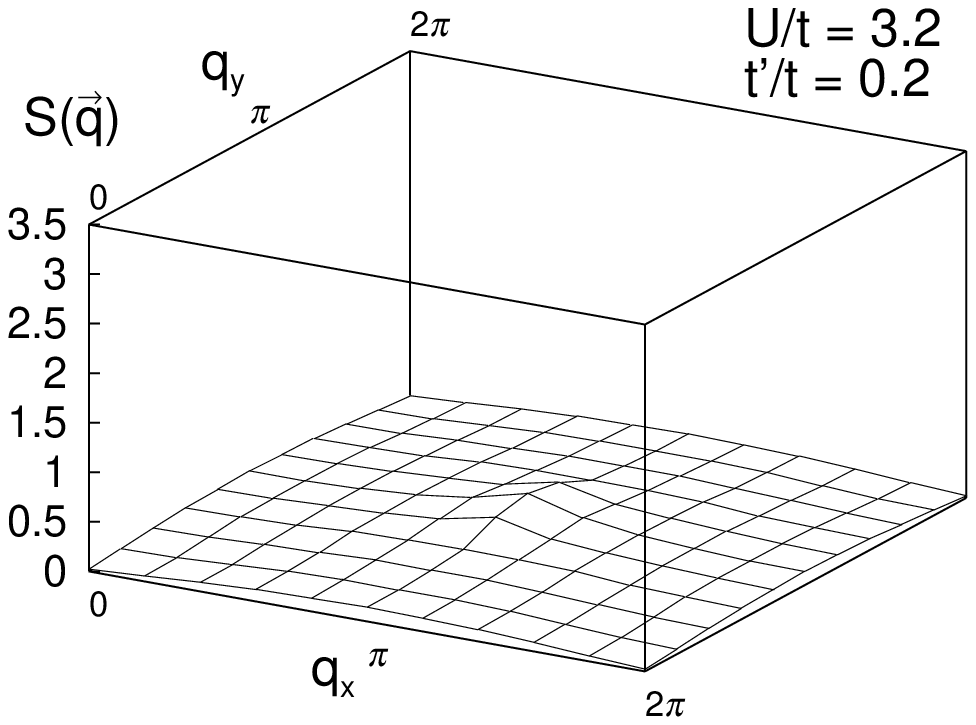}  \end{minipage}  
\hspace{3mm}  
\begin{minipage}{.55\linewidth}   \epsfxsize=80mm   \epsffile{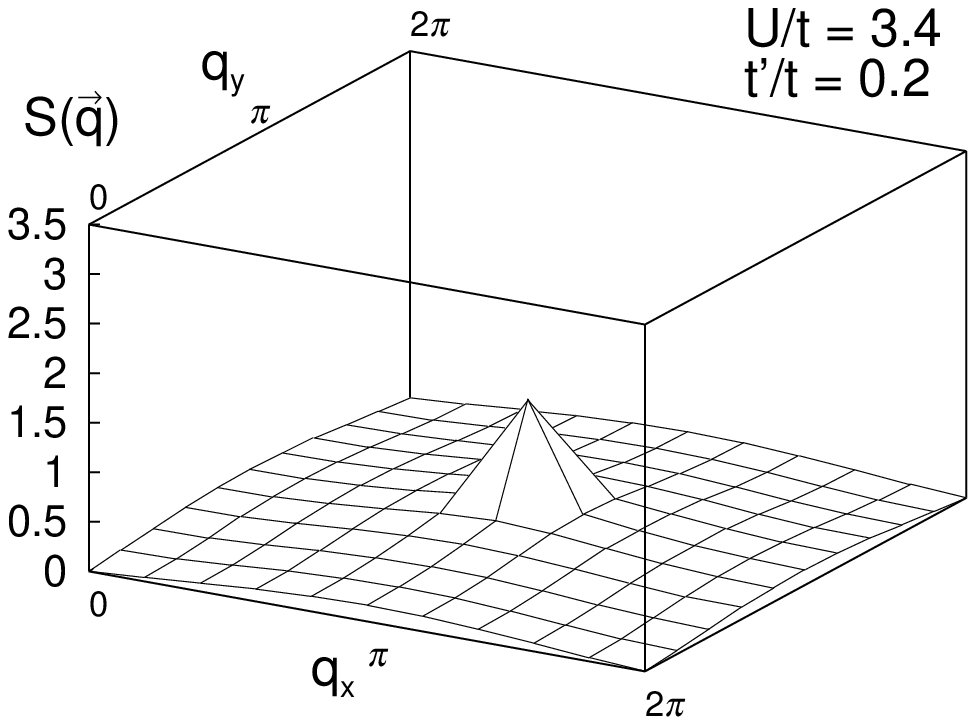}  \end{minipage}\\  
\begin{minipage}{.55\linewidth}   \epsfxsize=80mm   \epsffile{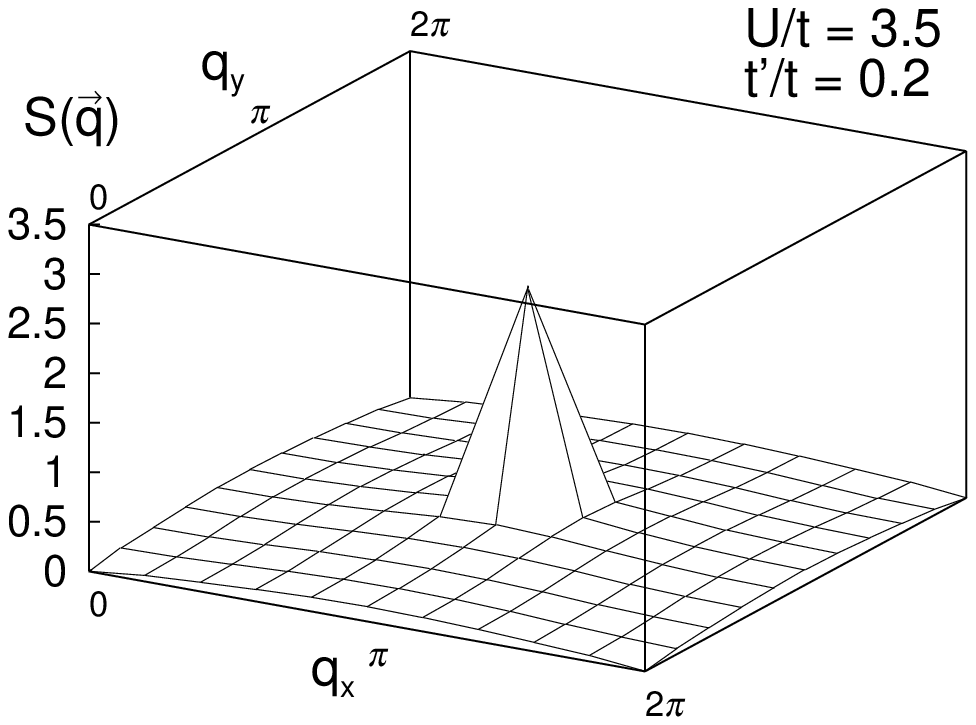}  
\end{minipage}  
\hspace{3mm}  
\begin{minipage}{.55\linewidth}   \epsfxsize=80mm   \epsffile{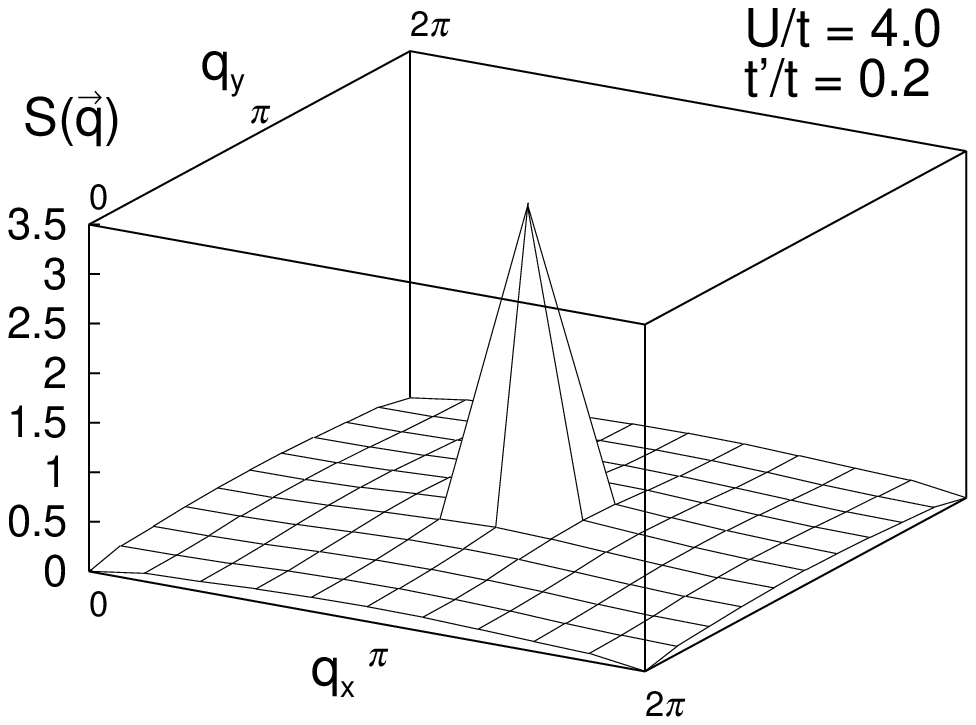}  
\end{minipage}  
\caption{The equal-time spin correlations in momentum space on $10\times 10$ 
 lattice at half filling for $t'/t=0.2$ and $U/t=3.2$, $3.4$, $3.5$ and $4.0$.}  
\label{spinstructure} 
\end{fullfigure} 

\subsection{$t'/t=0.5$ case}  
\subsubsection{Metal-insulator transition at $t'/t=0.5$} 
\begin{figure}  
\epsfxsize=80mm  
\epsffile{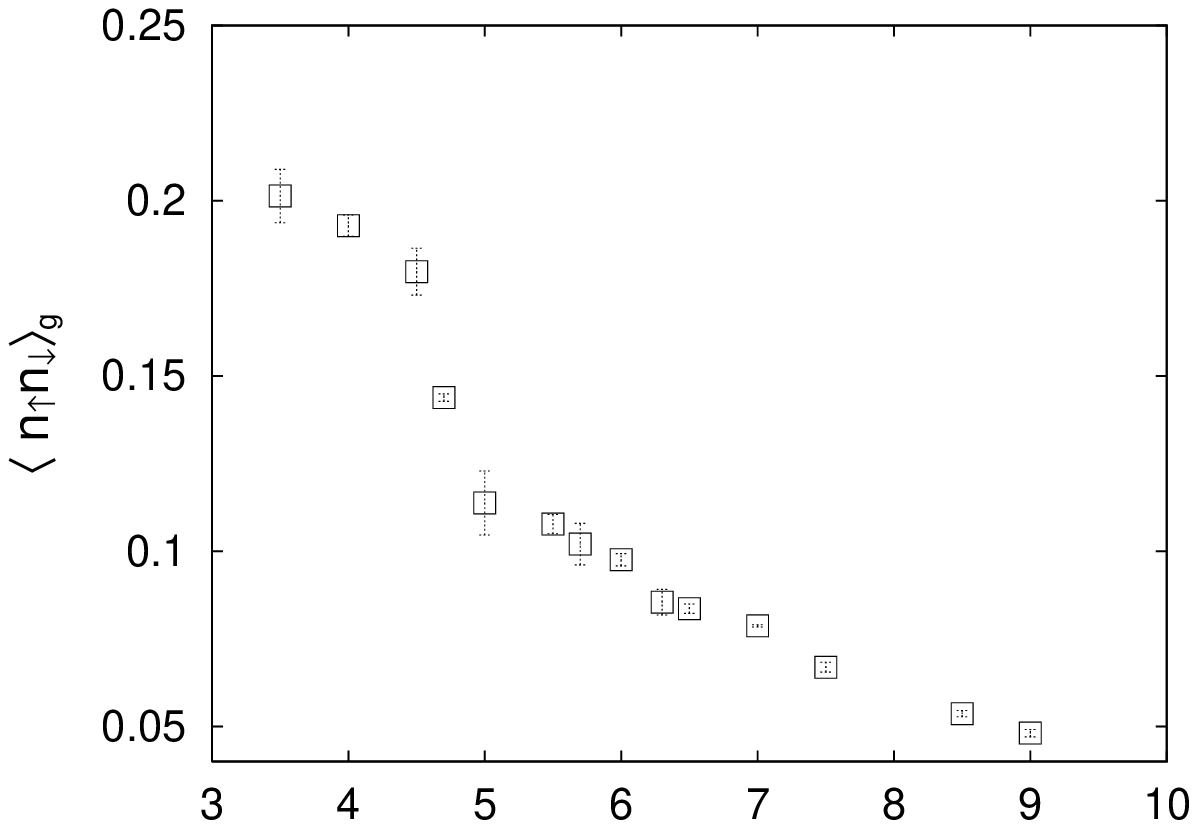}  
\caption{The expectation value of the double occupancy 
 $\langle n_{\uparrow}n_{\downarrow}\rangle_{g}$ after finite-size scaling. 
 The plots are for $t'/t=0.5$.} 
\label{DOB05} 
\end{figure}

Figure \ref{DOB05} shows the double occupancy after the size extrapolation 
as a function of the relative interaction $U/t$. A sharp change of the double occupancy occurs between $U/t=4.5$ and $5.0$. Differently from $t'/t=0.2$, this change does not seem to be a jump.  Hence, this transition at $U_{c1}/t=4.75\pm 0.25$ seems to be  a continuous transition.    

\begin{figure}  
\epsfxsize=80mm  
\epsffile{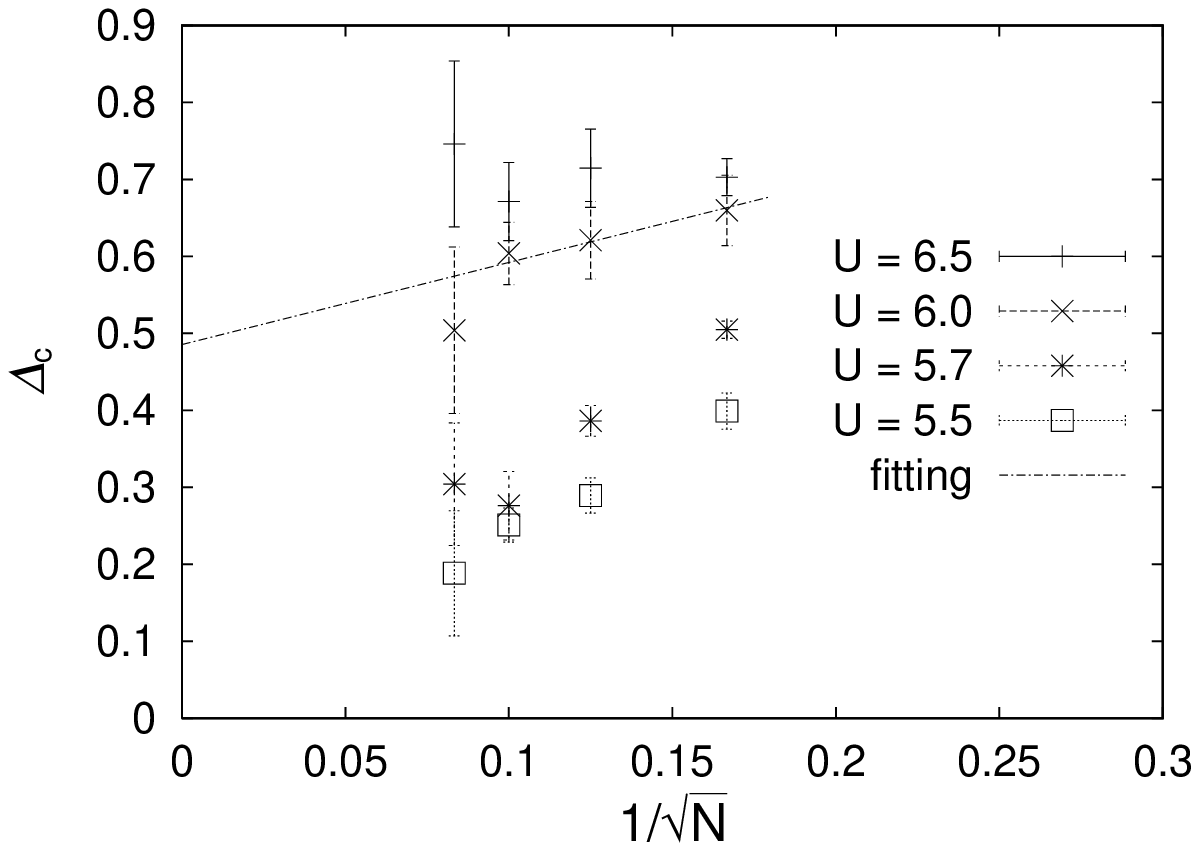}  
\caption{Finite-size scaling  of $\Delta_{c}$ for some choices of the  strength of the electron-electron interaction $U$ in $t'/t=0.5$ lattice. } 
\label{Fgap05} 
\end{figure}

Figure \ref{Fgap05} shows the charge excitation gap at larger $U$ than this transition point.  Although the finite size scaling is difficult, we can estimate the value of the charge excitation gap as in the same way as we have done at $t'/t=0.2$ using Eq.(\ref{gapEST}).  According to this equation, the charge excitation gap $\Delta_{c}$ at $U/t=6.0$ should be comparable or less than $(6.0-5.0)/2=0.5$ and $\Delta_{c}$ shown in Fig.\ref{Fgap05} satisfies this prediction.  Hence judging from the rough $U$ dependence of the charge gap,  we conclude that the transition seen in Fig.\ref{DOB05} is the metal-insulator transition.   

We also show the momentum distribution in Fig.\ref{momdis05}.  Although the qualitative change of the momentum distribution at $U/t=U_{c1}/t=4.75\pm 0.25$ can be seen,  it is remarkable that the change at the metal-insulator transition point is smaller and slower as a function of $U/t$ than the case at $t'/t=0.2$.  It may be related to the order of the transition; the fluctuation at the continuous transition is larger than that at the 
first-order transition and makes the change gradual.  The behavior of the momentum distribution at $U/t=4.5,t'/t=0.5$ implies that a strong renormalization of carriers may grow already in the metallic 
state
near $U_{c1}$ and an anomalous metal state may be stabilized. 
Spin fluctuations are not critically enhanced in this region because the 
spin correlations are short-ranged even in the insulating side near the phase boundary.  Therefore, unusual metallic behavior is not ascribed to the spin fluctuations.  The origin of the strong renormalization may come from solely charge fluctuations near the metal-insulator transition.
Superconducting pairing correlations are intriguing quantities to be studied in the future in terms of the instability of this unusual metal.
\begin{fullfigure}  
\begin{minipage}{.55\linewidth}   \epsfxsize=80mm   \epsffile{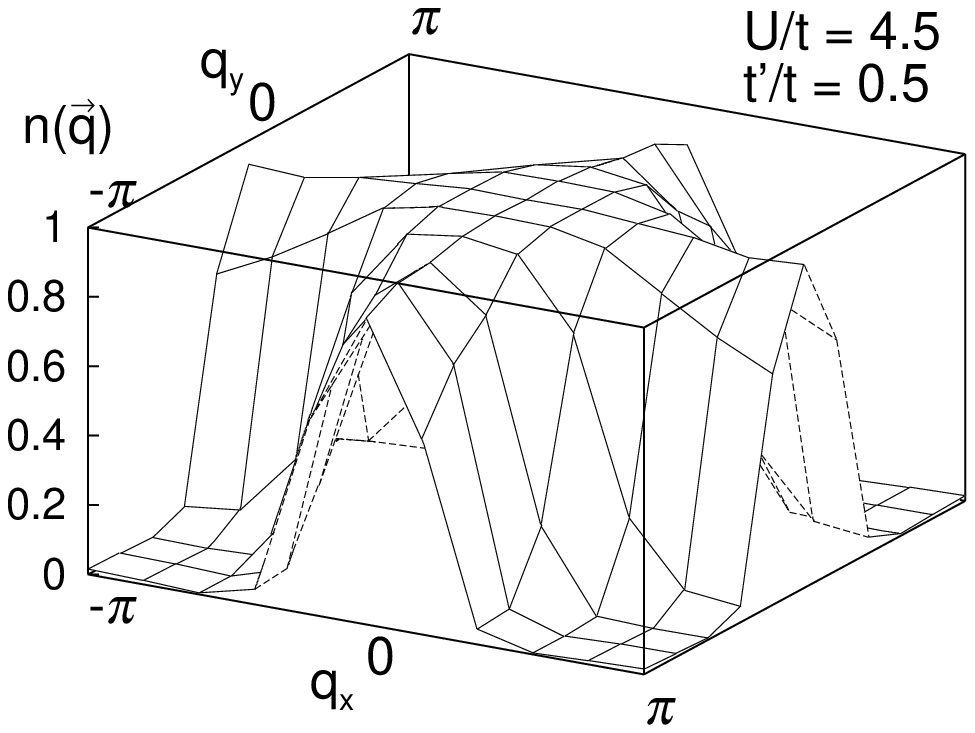}  
\end{minipage}  \hspace{3mm}  
\begin{minipage}{.55\linewidth}   \epsfxsize=80mm   \epsffile{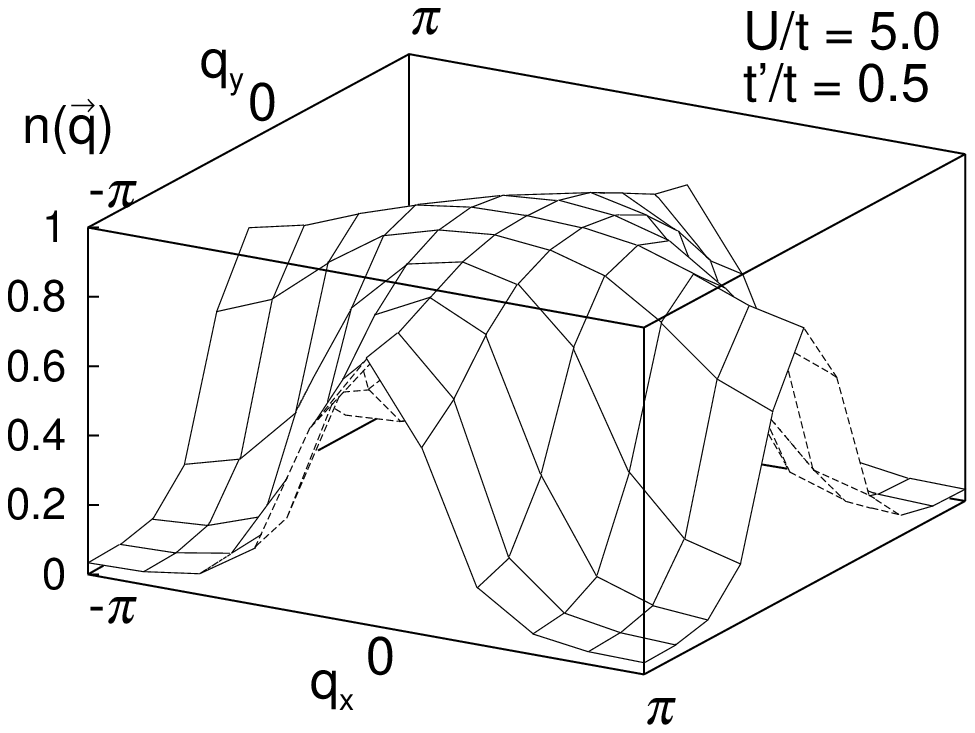}  
\end{minipage}\\  
\begin{minipage}{.55\linewidth}   \epsfxsize=80mm   \epsffile{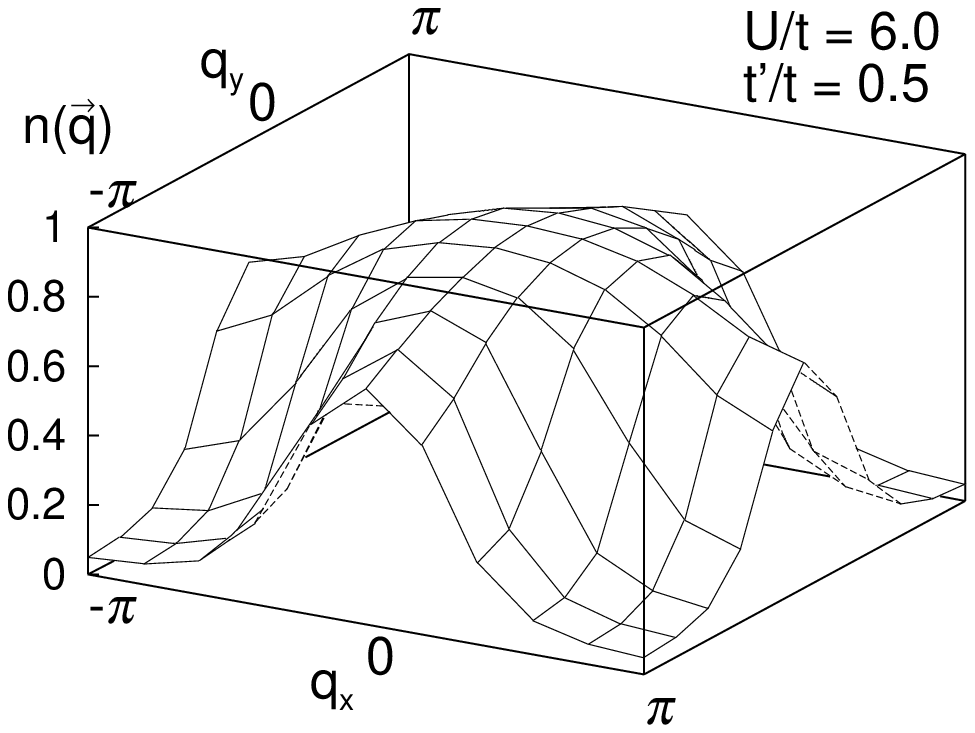}  
\end{minipage}  \hspace{3mm}  
\begin{minipage}{.55\linewidth}   \epsfxsize=80mm   \epsffile{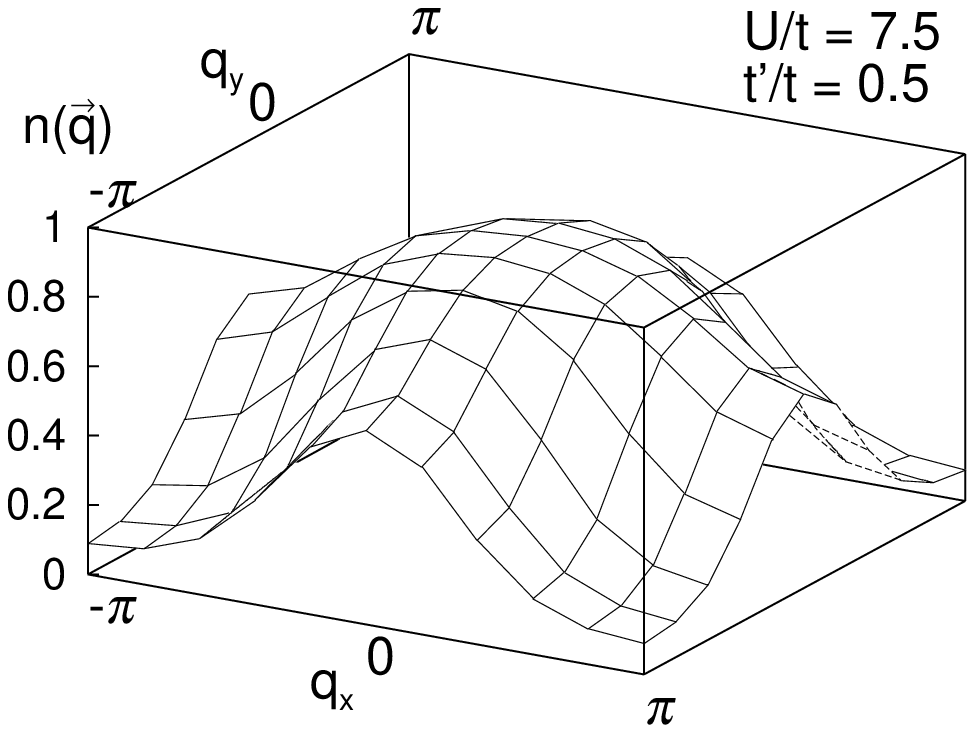}  
\end{minipage}  
\caption{The momentum distribution on $10\times 10$ lattice at  half filling for $t'/t=0.5$}  
\label{momdis05} 
\end{fullfigure}  

\subsubsection{Magnetic transition at $t'/t=0.5$} 
The structures of the equal-time spin correlations $S(\mib{q})$ are different from those for $t'/t=0.2$. Although the systems are at half filling,  Fig.\ref{spinstructure05} shows incommensurate peaks near $(\pi,\pi)$ for $U/t \le 7.0$.  The equal-time spin correlations have a single peak at $(\pi,\pi)$ for $6\times 6$ and $8\times 8$ lattices and have  incommensurate peaks for $10\times 10$ and $12\times12$ lattices.  We confirm that this tendency holds for $U/t=4.5$, $5.0$, $5.5$, $6.0$, $6.5$ and $7.0$.   

First we analyze the incommensurate peaks.  The incommensurate peaks shown in Fig.\ref{spinstructure05}  have asymmetric momentum dependence in terms of the four fold symmetry in $x$ or $y$ direction. These do not, however, have a physical significance of the symmetry breaking because the system size is finite.  The origin of this asymmetry may be that PIRG can treat only a subspace of the total Hilbert space and even after the extrapolation $L\rightarrow\infty$, this tendency remains.  We discuss this problem below.   

The similar incommensurate peaks appear away from half filling for the square lattice with $t'/t=0$\cite{QMC2}. In that case, the perfect nesting is destroyed by the hole doping and some other partial nesting occurs instead.  In the present case, however, the perfect nesting is destroyed by the frustrating term $t'$ and no clear nesting vectors at the Fermi level are found in the non-interacting system.   

\begin{figure}   
\epsfxsize=80mm   \epsffile{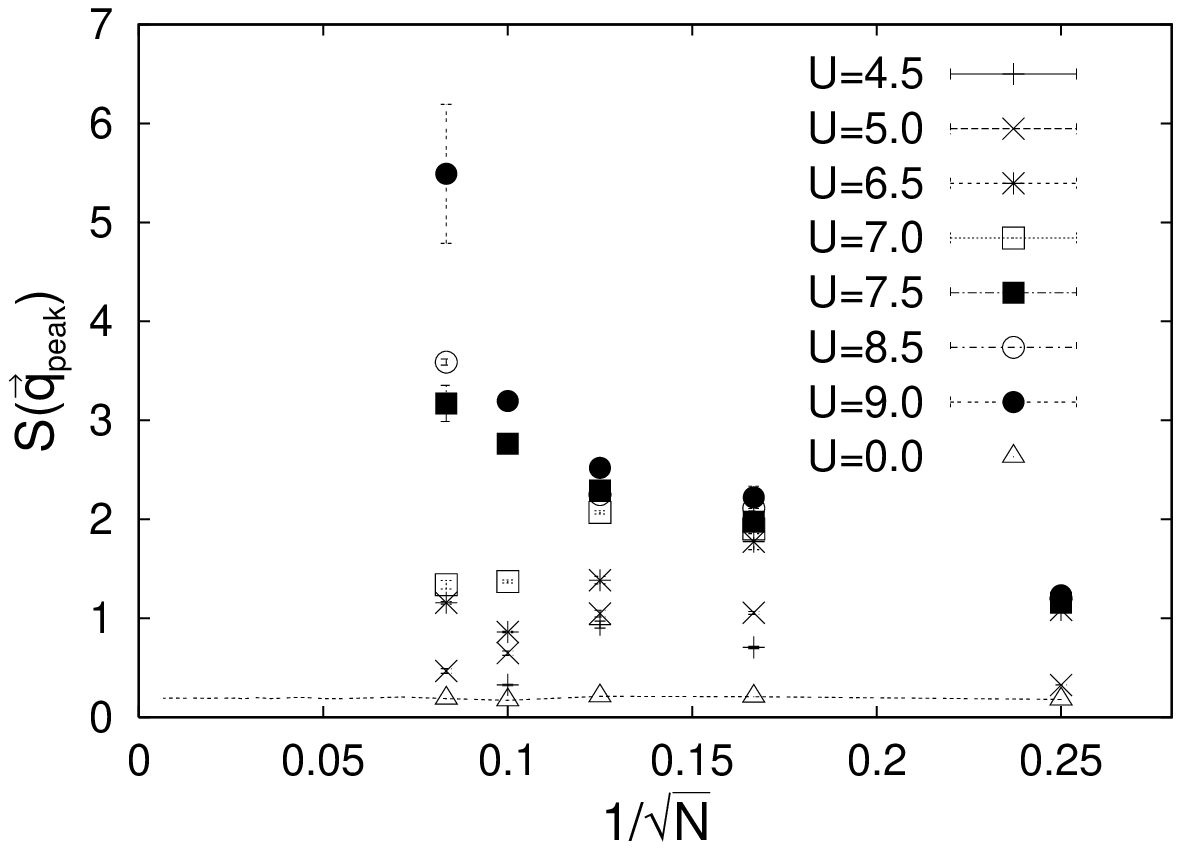}  
\caption{Finite-size scaling of $S\left(\mib{q}_{peak}\right)$ on some cases  of the strength of the electron-electron interaction $U$ in $t'/t=0.5$  lattice. }  
\label{peak05} 
\end{figure}

To discuss the presence of the long-range order, we have to study these superimposed incommensurate peaks in the thermodynamic limit.   In the following analysis, we assume that the occurrence of the incommensurate peaks is related to the flat structure of the peak at $(\pi,\pi)$ of the equal-time spin correlations for the non-interacting system as shown in Fig.\ref{spin05non}.  Though for $8\times 8$, $S(\pi,\pi)$ shows a single highest peak, for  $10\times 10$ and $12\times 12$, momenta $\mib{q}$ at which the $S(\mib{q})$ has  the highest value are $(\pi,\pi)$ as well as the nearest momenta to it.  This is a similar system-size dependence to that of the incommensurate peaks for the interacting system.  Although for $4\times 4, 6\times 6$ and $8\times 8$, there are single peaks at $(\pi,\pi)$, for $10\times 10$ and $12\times 12$, the incommensurate peaks as shown in Fig.\ref{spinstructure05} appear at the nearest momenta to $(\pi,\pi)$.  We also confirm that the `plateau' behavior holds for $30\times 30$ on non-interacting system.  Therefore the incommensurate peaks are likely to appear also at the nearest momenta to $(\pi,\pi)$ for the interacting systems larger than $12\times 12$ and the momenta at which peaks appear  can be speculated to converge to $(\pi,\pi)$ in the thermodynamic limit.  For this reason we just illustrate the peak values for $S(q)$
 on the same plane in Fig.\ref{peak05}. From this figure, it is clear that the behaviors of the equal-time spin correlations are different between $U\geq 7.5$ and $U\leq 7.0$.  Although the commensurate peaks of the equal-time spin correlations 
grow when the system size increases for $U\geq 7.5$, those for $U\leq 7.0$ remains finite clearly. We conclude that the incommensurate peaks which appear for $U\leq 7.0$ on the $10\times 10$ and $12\times 12$ lattices remain short-ranged in the thermodynamic limit and may converge to the commensurate peak.   

As we discuss in $\S$ \ref{SLIsec}, the asymmetry of the incommensurate peaks may come from a precursor of the four-sublittice order with Bragg peaks at $(\pi,0)$ and $(0,\pi)$.  When the short-ranged fluctuations around $(\pi,0)$ and $(0,\pi)$ occur in the Hilbert space far apart each other, PIRG may reproduce only one part of this fluctuation because of a limited number of $L$.  Since PIRG may be regarded as a systematic improvement of the Hartree-Fock approximation, and the Hartree-Fock calculation
always overestimate the ordered phase, 
the absence of the long-ranged order observed by PIRG should be reliabe for this incommensurate fluctuations.   

\begin{figure}  
\epsfxsize=80mm  \epsffile{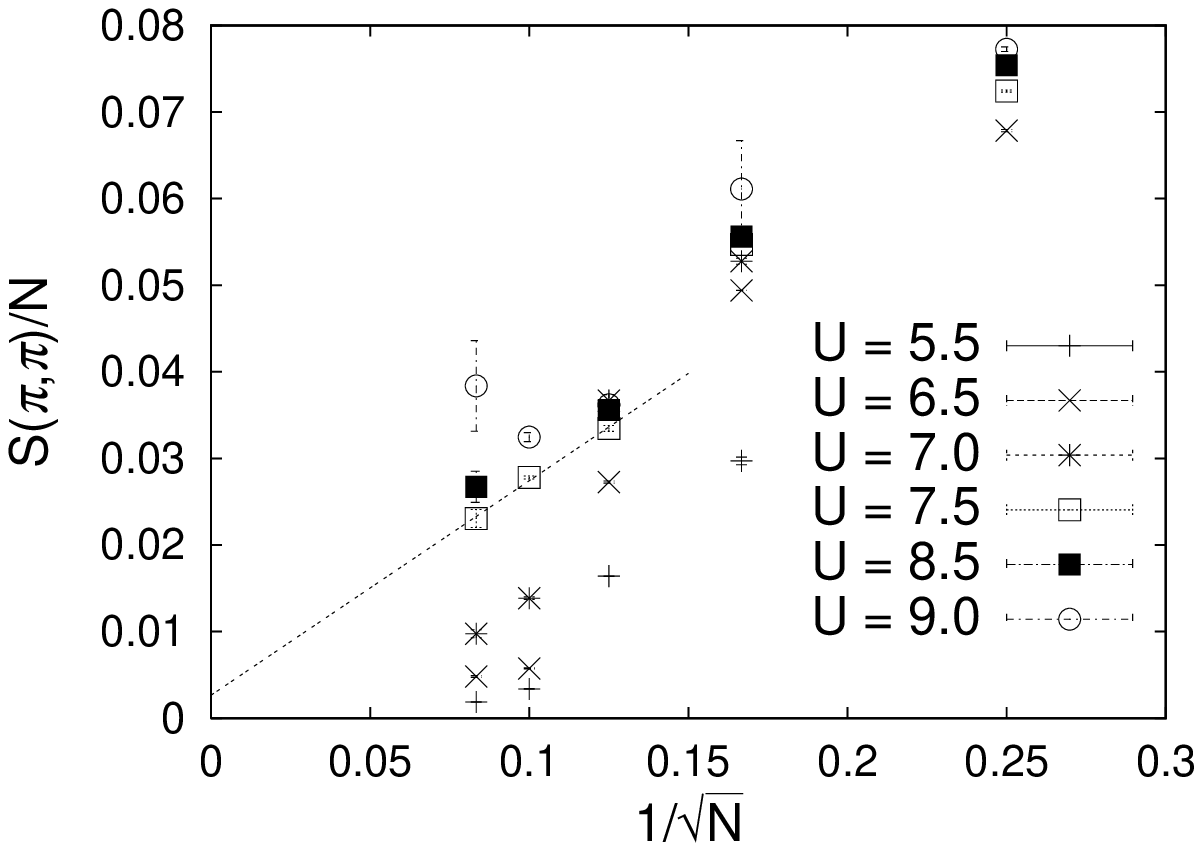}  
\caption{Finite-size scaling of $S\left(\mib{q}_{peak}\right)$ on some cases  of the strength of the electron-electron interaction $U$ at $t'/t=0.5$. }  
\label{H05} 
\end{figure}

Next we discuss the commensurate long-range order.    The incommensurate peaks disappear and  the commensurate antiferromagnetic peak at $(\pi,\pi)$ appears for  $U\geq 7.5$ as we see in Fig.\ref{spinstructure05}.  To discuss the long-range order, we use the finite-size scaling function  Eq.(\ref{FSS}) again and Fig.\ref{H05} shows the result.  This figure shows that the commensurate peak which appears for $U\geq 7.5$ indicates the long-ranged order.  The transition bears continuous character.  The reason for that we can not detect a jump of $\partial E_{g}/\partial U$ is supposed to be the same as the reason for the case $t'/t=0.2$; the jump is too small.  
\begin{fullfigure}  
\begin{minipage}{.55\linewidth}   \epsfxsize=80mm   \epsffile{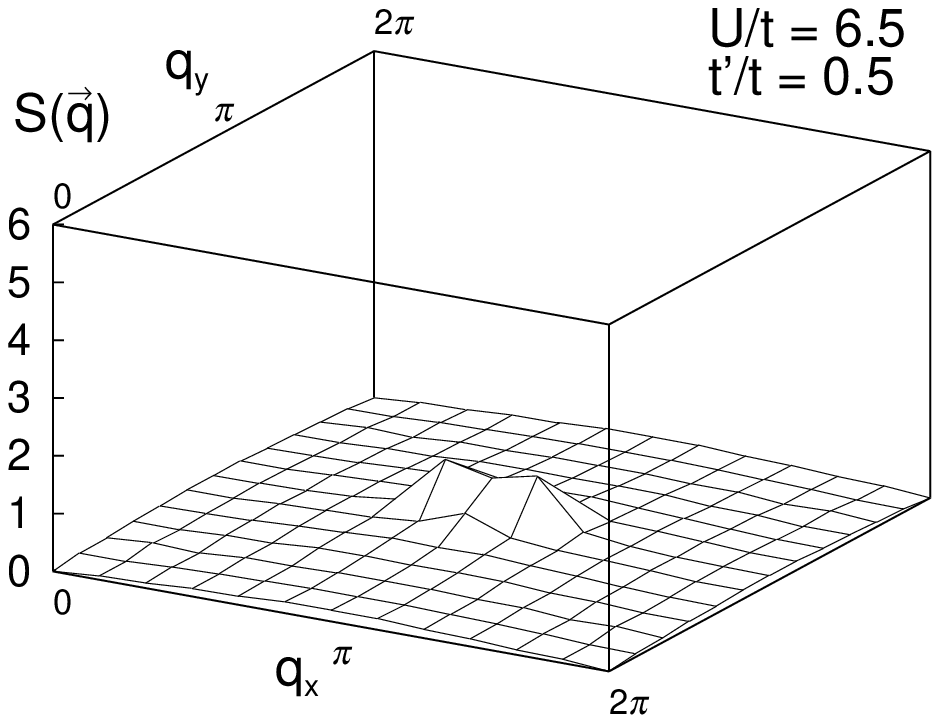}  
\end{minipage}  \hspace{3mm}  
\begin{minipage}{.55\linewidth}   \epsfxsize=80mm   \epsffile{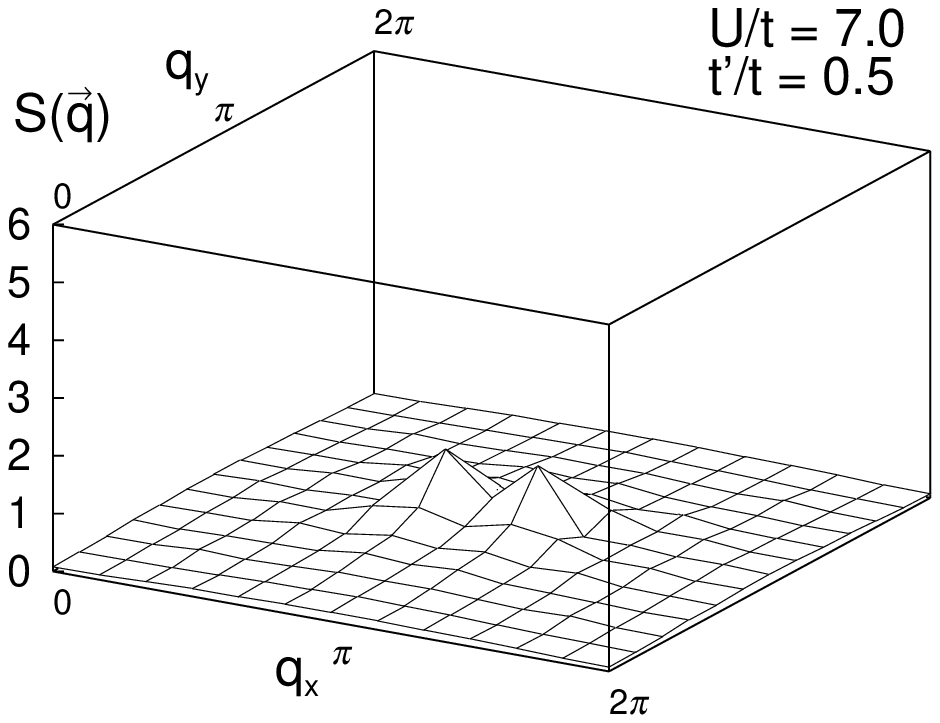}  
\end{minipage}\\  
\begin{minipage}{.55\linewidth}   \epsfxsize=80mm   \epsffile{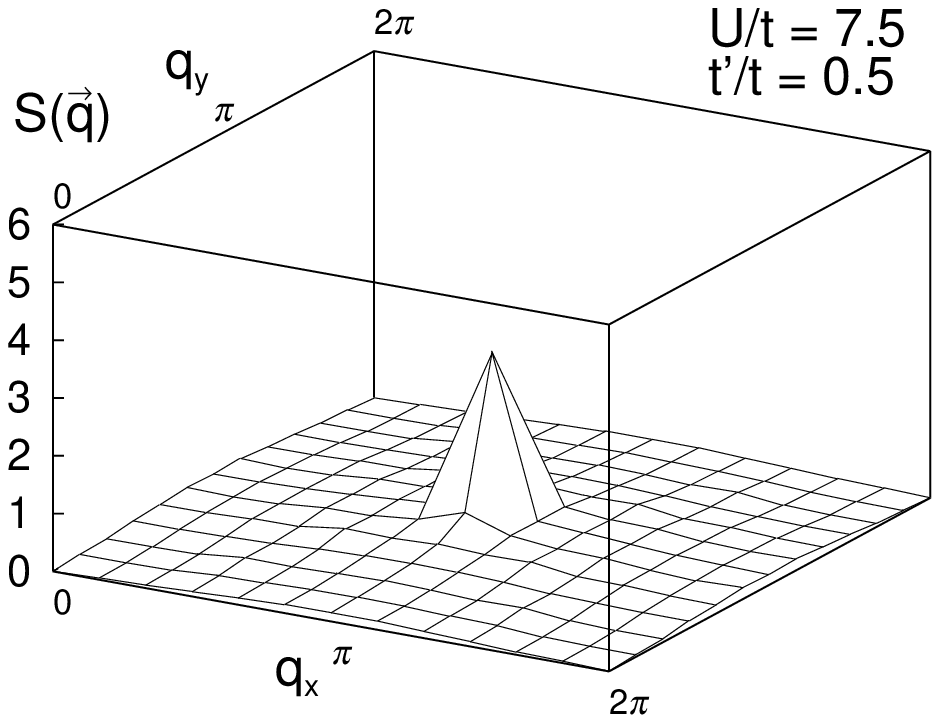}  
\end{minipage}  \hspace{3mm}  
\begin{minipage}{.55\linewidth}   \epsfxsize=80mm   \epsffile{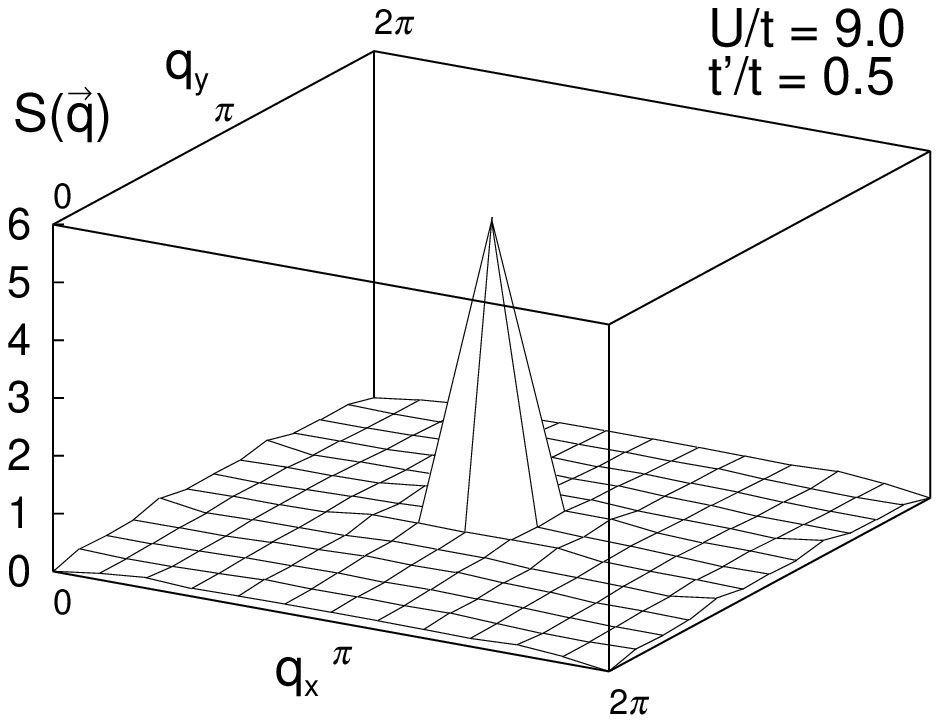}  
\end{minipage}  
\caption{The equal-time spin correlations in the momentum space on  $12\times 12$ and $t'/t=0.5$ lattice at half filling for $U=6.5$, $7.0$,  $7.5$ and $9.0$. Note that the asymmetry is caused by the PIRG.}   
\label{spinstructure05} 
\end{fullfigure} 
\begin{fullfigure}  
\begin{minipage}{.55\linewidth}   \epsfxsize=80mm   \epsffile{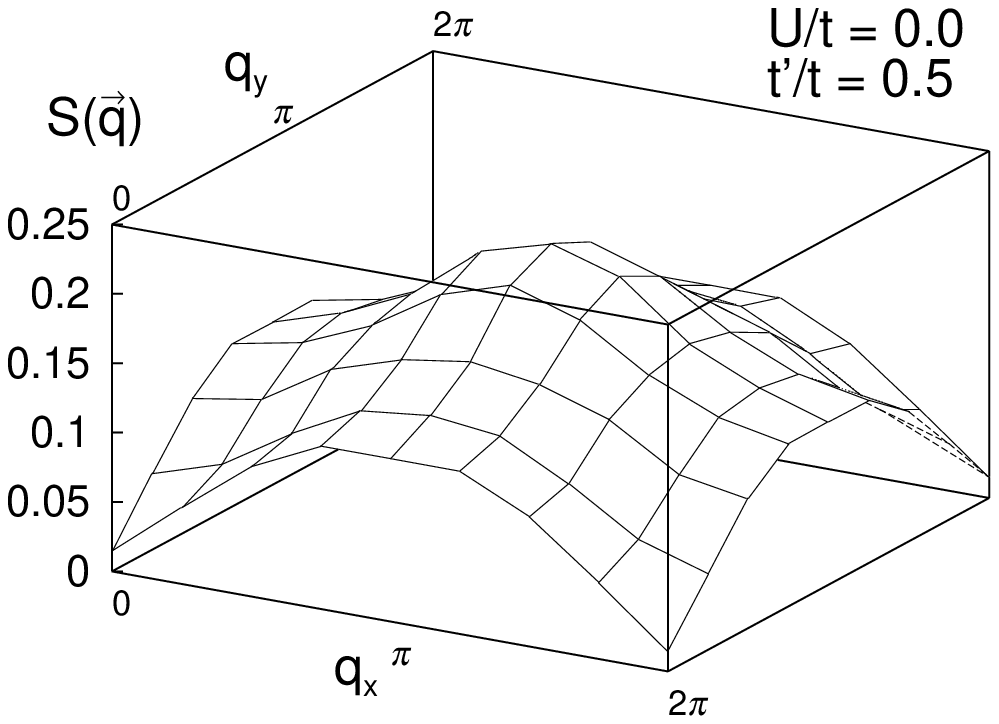}  
\end{minipage}  \hspace{3mm}  
\begin{minipage}{.55\linewidth}   \epsfxsize=80mm   \epsffile{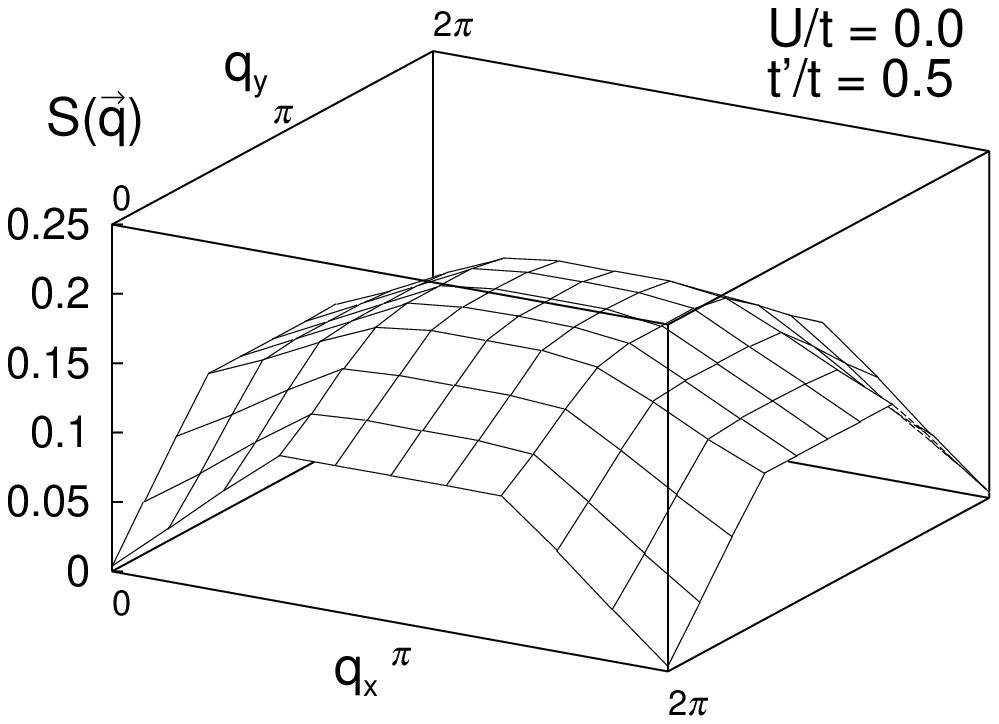}  
\end{minipage}\\  
\begin{minipage}{.55\linewidth}   \epsfxsize=80mm   \epsffile{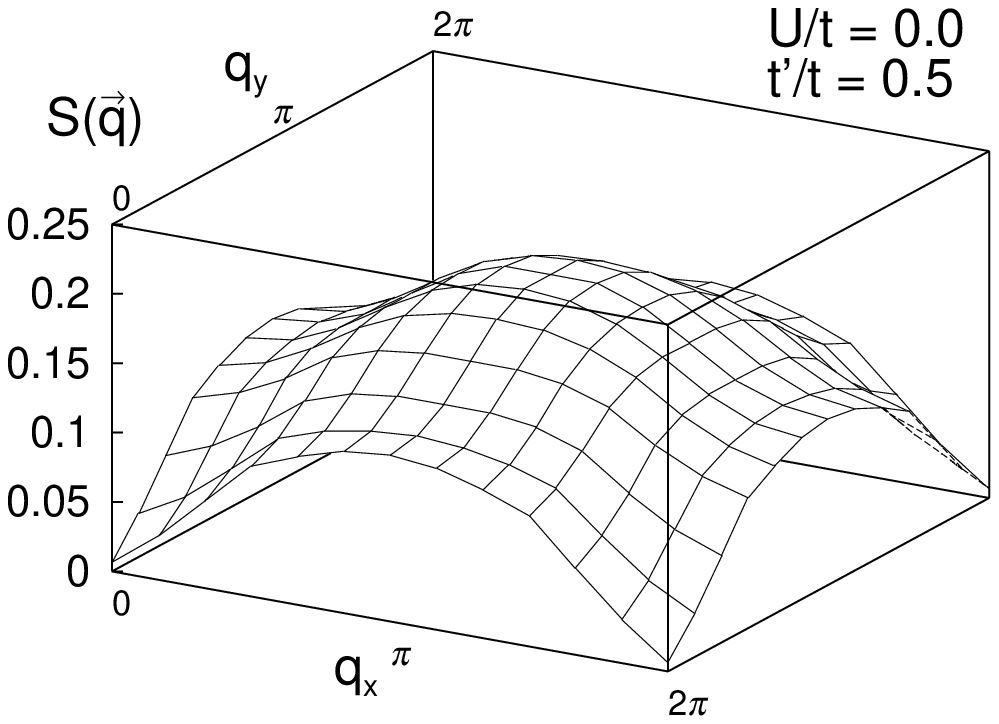}  
\end{minipage}  \hspace{3mm}  
\begin{minipage}{.55\linewidth}   \epsfxsize=80mm   \epsffile{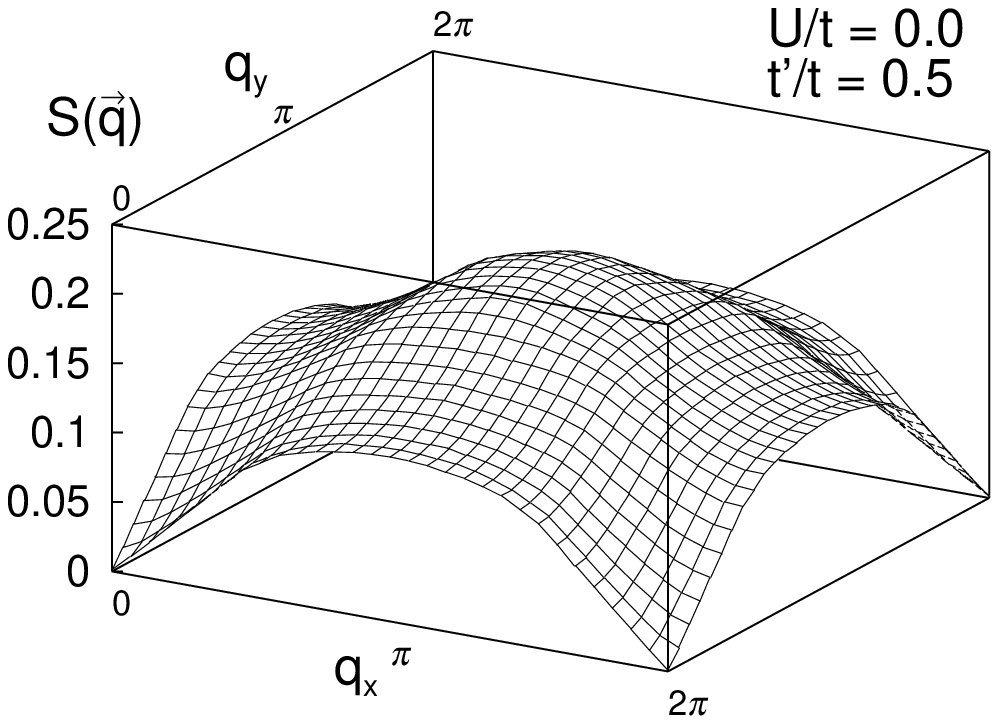}  
\end{minipage}  
\caption{The equal-time spin correlations in the momentum space for $t'/t=0.5$  at half filling for the non-interacting system. The systems  are $8\times 8$, $10\times 10$, $12\times 12$ and $30\times 30$ lattices.
 For lattices equal to or larger than $10\times 10$, the peak always has 
 five-fold degeneracy around $(\pi,\pi)$ which leads to the `plateau' structure
 around $(\pi,\pi)$. }  
\label{spin05non} 
\end{fullfigure} 
\begin{figure}  
\epsfxsize=80mm  \epsffile{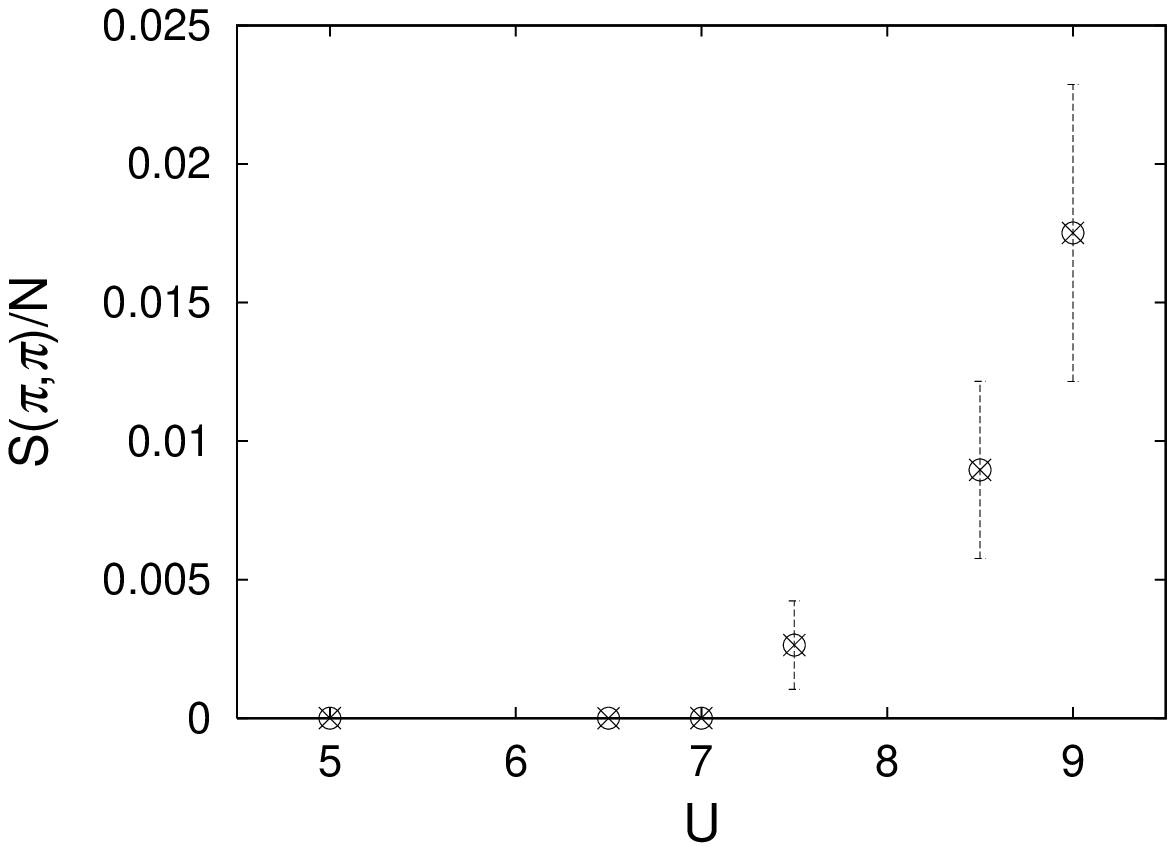}  
\caption{$S\left(\pi,\pi\right)/N$ in the thermodynamic limit as a  function of the strength of the electron-electron interaction $U$ in  half-filled systems at $t'/t=0.5$. }  
\label{mag05} 
\end{figure} 
\begin{table}  
\caption{Critical value $U_{c}/t$ for magnetic transitions to a  commensurate antiferromagnetic state and metal-insulator transitions at  half filling for $t'/t=0.2$. }  
\begin{center}  
\begin{tabular}{@{\hspace{\tabcolsep}\extracolsep{\fill}}ccc}   \hline    & magnetic $U_{c2}/t$ & metal-insulator $U_{c1}/t$ \\   \hline   PIRG & $3.45\pm 0.05$ & $3.25\pm 0.05$ \\   Hartree Fock \cite{HF} & $2.064$ & $2.064$ \\   Green's function \cite{HF2} & $4.00$ & $4.00$ \\   Gutzwiller \cite{HF}& $3.920$ & no data \\   Quantum Monte Carlo \cite{Hirsch} & $2.5\pm 0.25$ & no data \\   \hline  
\end{tabular}   
\end{center}  
\label{table02} 
\end{table} 
\begin{table}  
\caption{Critical value $U_{c}/t$ for magnetic transitions to a  commensurate antiferromagnetic state and metal-insulator transitions at  half filling for $t'/t=0.5$. }  
\begin{center}   
\begin{tabular}{@{\hspace{\tabcolsep}\extracolsep{\fill}}ccc}    \hline    & magnetic & metal-insulator \\    \hline    PIRG & $7.25\pm 0.25$ & $4.75\pm 0.25$ \\    Hartree Fock \cite{HF} & $3.215$ & $3.290$ \\    Gutzwiller \cite{HF} & $5.259$ & no data \\    \hline   
\end{tabular}  
\end{center}  
\label{table05} 
\end{table} 

\subsection{Comparison with Hartree-Fock results\cite{HF}} 
The critical values of the relative interaction $U_{c}/t$ determined above is shown in Tables \ref{table02} and \ref{table05}.  The magnetic transition appears to occur at larger $U$ than that for the metal-insulator transition. The paramagnetic insulating state seems to appear in between. This is in contrast with the Hartree-Fock calculation where the two transitions occur concurrently at $t'/t=0.2$ while 
the two transitions occur in the opposite order and an antiferromagnetic metal is stabilized at $t'/t=0.5$.  This may be interpreted from the fact that the magnetic long-ranged order is more strongly overestimated in the Hartree-Fock approximation than the metal-insulator transition because fluctuation effects are more serious for the antiferromagnetic symmetry breaking. 

The Hartree-Fock approximation also shows that the combined transition at $t'/t=0.2$ is of the first order\cite{HF}.  PIRG results appear to show that the two transitions are both of first order at $t'/t=0.2$.  At $t'/t=0.5$,  the Hartree-Fock approximation study claims that the magnetic transition is of second order and the PIRG results does not seem to contradict this.  PIRG also shows that the metal-insulator transition seems to become continuous.    

\subsection{Presence of spin liquid insulator near the   metal-insulator phase boundary} 
\label{SLIsec} 
According to the above discussion, there is a nonmagnetic insulator  phase between metal-insulator and magnetic transitions.  We discuss this remarkable fact in detail for the case of $t'/t=0.5$.  There have been not many studies on the Hubbard model on a square lattice with nearest-neighbor and next-nearest-neighbor transfers. However, extensive studies have been done on spin-$1/2$ Heisenberg model on the same lattice\cite{firstliquid,Gelfand,Nakamura,refspinliquid,SL_CO,CO}  which is called $J_{1}$-$J_{2}$ model, although the result is not sufficiently conclusive due to the negative sign problem and lack of efficient numerical algorithm.   The Hamiltonian of the $J_{1}$-$J_{2}$ model is  
\begin{eqnarray}  
\hat{H}&=&J_{ij}\sum_{ij}\bf{S}_{i}\cdot\bf{S}_{j}\nonumber\\  J_{ij}&=&   \left\{    \begin{array}{lll}     J_{1}>0 \textrm{ for $i$ and $j$ are the nearest neighbor sites}\\     J_{2}>0 \textrm{ for $i$ and $j$ are the next nearest neighbor sites}\\     0 \textrm{    otherwise}    \end{array}     \right.
 \label{Heisenberg} 
\end{eqnarray} 
where $\bf{S}_{i}$ is the quantum spin operator ($S=\frac{1}{2}$) of the  $i$-th lattice site. The ground state of the classical $J_{1}$-$J_{2}$ model on the square lattice is the two-sublattice N\'{e}el state for $J_{2}/J_{1}<0.5$ and the degenerate four-sublattice N\'{e}el state for $J_{2}/J_{1}>0.5$. Similarly to the Hubbard models, the  quantum model with $J_{1}$ only has the N\'{e}el antiferromagnetic long-range order in the ground state. According to the literature\cite{SL_CO,CO}, increasing the frustration $J_{2}$  destabilizes the N\'{e}el order and at $J_{2}/J_{1}\simeq 0.4$ a phase transition is inferred to occur to some different state and at $J_{2}/J_{1}\simeq 0.6$ a phase transition to a  collinear order, which is a special case of the four-sublattice N\'{e}el state, occurs. Although no conclusive results are available, the state stabilized for $0.4<J_{2}/J_{1}<0.6$ is speculated to be a nonmagnetic (or spin liquid) insulator.   In fact, a translational symmetry breaking due to a dimer order has been suggested\cite{Sachdev,Sushkov}, although it is controversial\cite{Sorella2}.     

It is known that the half-filled Hubbard model can be treated by the second-order perturbation and in the limit of large interactions, the Hubbard model is mapped to the Heisenberg model with the superexchange coupling,  
\begin{equation}  
J_{ij}=\frac{4|t_{ij}|^{2}}{U}. 
\end{equation} 
Now we focus on the case $t'/t=0.5t$. This parameter value leads to the mapping:  
\begin{equation} 
t'/t=0.5\quad\Longrightarrow\quad J_{2}/J_{1}\simeq 0.25. 
\end{equation} 
In this region, the ground state of the $J_{1}$-$J_{2}$ model seems to have an antiferromagnetic long-range order of the N\'{e}el state.  This is also supported by our result for $U/t\geq 7.5$.  According to our  result on the Hubbard model, however, there is an insulating phase without the antiferromagnetic long-range order for $5.0\leq U/t\leq 7.5$.  The reason for the appearance of this intermediate state for $5.0\leq U/t\leq 7.5$ in the Hubbard model is speculated in the following: In the Heisenberg model, the driving force to destabilize the commensurate antiferromagnetic long-range order is the interplay of quantum mechanical fluctuations and the frustration $J_{2}$.  However, in addition to this interplay, charge fluctuations ascribed to a finite amplitude of double occupancy reduces the magnetic long-range order in the Hubbard model  at finite $U$ and enhances quantum fluctuations.  Then it is likely that the  nonmagnetic insulator may be stabilized even for $t'$ smaller than $0.63\left(\Leftrightarrow J_{2}/J_{1}\simeq 0.4\right)$ if $U$ decreases from infinity.  Moreover, incommensurate peaks and their symmetry breakings to $x$ or to $y$ direction may be understood in the following:  $S(\mib{q})$ of two-sublattice N\'{e}el state stabilized at $J_{2}/J_{1}<0.4$ has a Bragg peak at $\mib{q}=(\pi,\pi)$ and $S(\mib{q})$ of four-sublattice collinear state stabilized at $J_{2}/J_{1}>0.6$ has a Bragg peak at $\mib{q}=(\pi,0)$ or $(0,\pi)$. The dimer order or its fluctuations may also induce a peak around $\mib{q}=(\pi,0)$ or $(0,\pi)$. Hence,  $S(\mib{q})$ of a nonmagnetic insulator state, stabilized in the intermediate region, can be speculated to have peaks between $(\pi,\pi)$ and $(\pi,0)$, or between $(\pi,\pi)$ and $(0,\pi)$. Incommensurate peaks shown in Fig.\ref{spinstructure05} is indeed found in this momentum region.  Though these incommensurate peaks may appear only for finite-size systems, the physics contained in it may be related to a precursor and fluctuations of the collinear or the dimer state.  Even when the dimer or plaquet orderings leading to the translational symmetry breakings is stabilized in the Heisenberg limit 
$U \rightarrow \infty$, such translational symmetry breakings would be destroyed under sufficient charge fluctuations realized with reducing $U$ near the metal-insulator transition.  The nature of the nonmagnetic insulator phase found between the magnetic and metal-insulator transitions is an outstanding and fundamental problem to be clarified.  This interesting issue will be studied in separate work.
\subsection{Phase diagram} 
\label{Phasesec} 
\begin{figure}  
\epsfxsize=80mm  \epsffile{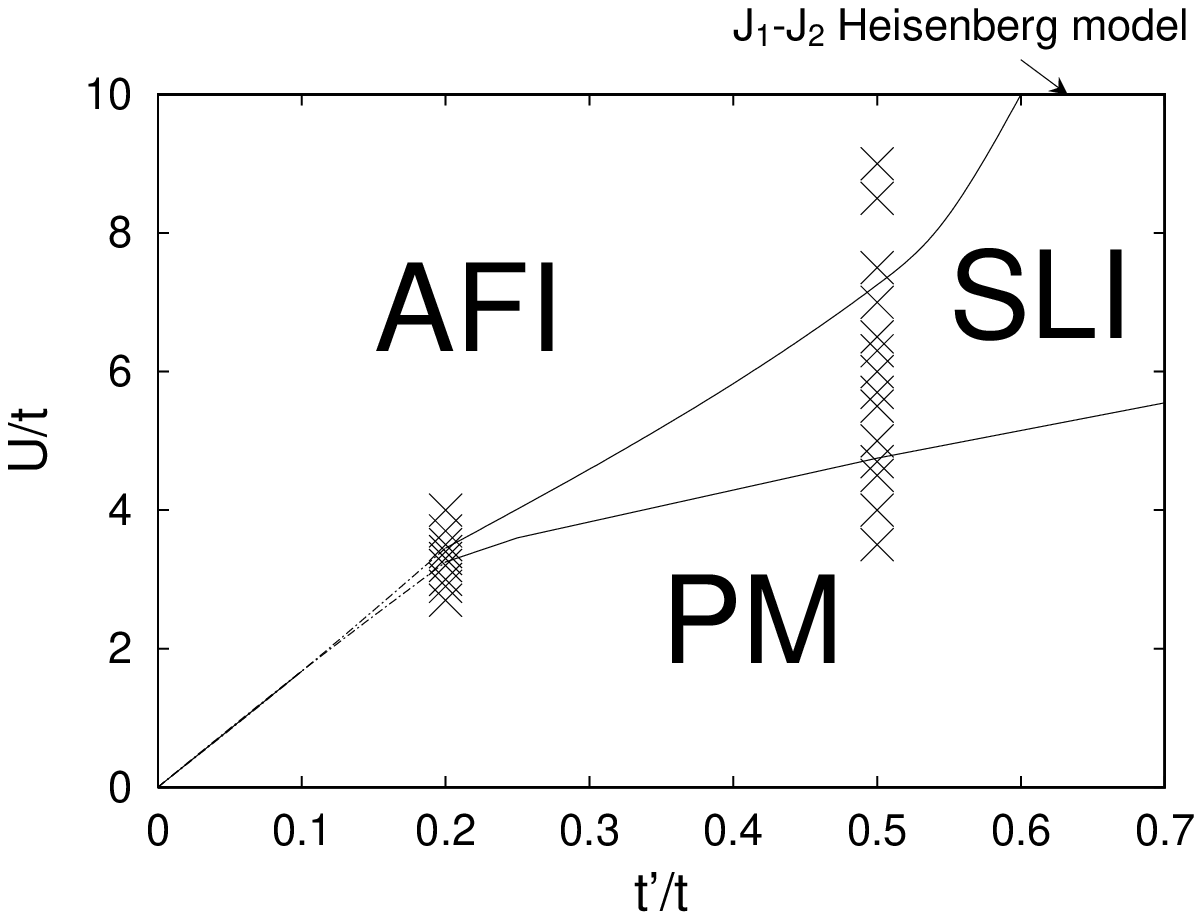}  
\caption{A schematic phase diagram in the plane of $U/t$ and $t'/t$. 
 The symbols $\times$ represent  points where  we have analyzed in detail by the PIRG calculation. 
 Antiferromagnetic insulator, spin liquid insulator (or nonmagnetic insulator), 
 and paramagnetic metal are denoted by AFI, SLI, and PM,
 respectively.  The arrow indicates the phase boundary between AFI 
 and SLI in the limit $U/t \rightarrow \infty$ roughly speculated 
 from the frustrated Heisenberg model.}  
\label{phdiag} 
\end{figure}

By using the PIRG results, the result of the frustrated Heisenberg model,
 and the fact  that the system is in a paramagnetic-metal phase at $U/t=0$ while in an antiferromagnetic-insulator phase at $t'/t=0$, we show a schematic phase diagram in Fig.\ref{phdiag}.   
For smaller $t'/t$ up to 0.2, the metal-insulator and antiferromagnetic transitions occur more or less concurrently with a first-order nature similarly
to the Hartree-Fock results.  On the other hand, with increasing $t'/t$, the phase diagram shows qualitative difference from the Hartree-Fock prediction. The appearance of the nonmagnetic insulator rather has a consistency with the strong coupling Heisenberg picture.  The present numerical results clarify how these two contradicting pictures are reconciled at finite $U$.
The emergence of unusual metallic behavior observed near the phase boundary between the nonmagnetic insulator and the paramagnetic metal is one of the most remarkable findings in this phase diagram. 
We have not studied the spin excitation gap. This is left for future studies. 
 
In the phase diagram, we discuss the regimes of the organic materials and the cuprate superconductors.  Although the organic materials such as BEDT-TTF compounds have different lattice structure from the present $t$-$t'$-$U$ model, these compounds seem to be located in the regime of smaller $U/t$ and $t'/t$ near the concurrent transition boundary, because they show concurrent trantions in many cases, while the nonmagnetic insulator except the spin-Peierls phase has not been observed so far.  The cuprate superconductors may be mimicked by the present model while they undergo the filling-control transition contrary to the present bandwidth-control study.  In any case, these superconductors with higher transition temperatures such as Bi, Tl, and Hg compounds seem to have the effective $t'/t \sim 0.3-0.4$ with $U/t \sim 8$\cite{Andersen}.  This is deeply inside the AFI phase at half filling, namely in the mother materials.

\section{Summary and Discussion} 
\label{conc} 
Mott transitions in strongly correlated two-dimensional systems may occur through various different routes such as bandwidth control, band filling control and dimensionality control.  The filling-control transition has already been studied quite extensively in two dimensions\cite{QMC1,QMC2,MIT}.  Because the Mott transition in the half-filled square lattice with the nearest-neighbor transfer is expected to occur at $U=0$ owing to the perfect nesting, it is necessary to treat the systems with longer-ranged transfer for studying the bandwidth-control transition.  Before PIRG, no powerful numerical methods had been available except in  one-dimensional systems for discussing such a system because of the sign problem in the Monte Carlo method and for this reason the bandwidth-control 
transition
has not been studied on the same level as the filling control case.  In this paper, by applying PIRG, we have for the first time studied 
the bandwidth-control transition systematically.   We have studied the two-dimensional Hubbard model on the square lattice with next-nearest-neighbor transfers $t'$ to clarify the nature of the bandwidth-control transition.  More concretely, we determine the critical strengths of the on-site Coulomb interaction $U_{c}/t$ of the  magnetic transition and of the  metal-insulator transition for two choices of the parameters  $t'/t = 0.2$ and $0.5$ for the Hamiltonian (\ref{Hamiltonian}).  For these systems, there exist some results from the Hartree-Fock approximation in the literature, although they may be uncontrolled owing  to the neglect of fluctuations.  As easily expected, because of the fluctuations, the critical strengths of the interactions estimated by PIRG are larger than those estimated by the Hartree-Fock approximation on metal-insulator and magnetic transitions.  The present PIRG results show that the metal-insulator transition occurs at $U_{c1}$ which is lower than the antiferromagnetic transition point $U_{c2}$, in sharp contrast with the Hartree-Fock prediction. The Hartree-Fock result gives the antiferromagnetic transition at lower $U$ than the metal-insulator transition and an antiferromagnetic metal appears in between the two transitions. On the contrary, in the region $U_{c1}<U<U_{c2}$, the nonmagnetic insulating state appears to be stabilized in our results.   

For the case $t'/t=0.2$, the critical value of the metal-insulator transition is $U_{c1}/t=3.25\pm 0.05$ and that of the magnetic transition is $U_{c2}=3.45\pm 0.05$.  It appears to show two separate transitions in our accuracy.  We have analyzed the order of the magnetic transition for $t'/t=0.2$.  $S(\pi,\pi)$ appears to diverge suddenly between $U/t=3.4$ and $U/t=3.5$ without appreciable growth of short-ranged correlations for $U<3.4$.  The transition is consistent with that of the first order as a function of the electron-electron interaction $U$ as the same as the result of the Hartree-Fock approximation.  The staggered magnetization $m$ in the thermodynamic limit seems to have a sharp increase  at the critical strength of the electron-electron interaction in Fig.\ref{U_S} and it does not contradict the above conclusion, although a sharp continuous transition is not excluded. We have also analyzed the order of metal-insulator transition. It also appears to show a first-order transition with a jump in the average of the double occupation.    

For the case $t'/t=0.5$, the metal-insulator transition occurs at $U_{c1}/t=4.75\pm 0.25$ and the magnetic transition does at $U_{c2}/t=7.25\pm 0.25$.  The difference between two transition points, $|U_{c1}-U_{c2}|$,
 become wider with increasing $t'$.  Incommensurate peaks of the equal-time spin correlation appear near $(\pi,\pi)$ in finite-size systems for $U/t\leq 7.0$.  We have analyzed these incommensurate peaks by analyzing the amplitude   of the highest peaks.  This analysis indicates that these incommensurate peaks do not have a long-range order.  $S(\pi,\pi)$ between $U/t=7.0$ and $U/t=7.5$ suggests that the magnetic transition is of more 
continuous type. The staggered magnetization also supports this.  Figure \ref{DOB05} suggests that the average of double occupation change continuously at the transition point and metal-insulator transition 
also behaves rather as a continuous one.  The behavior of the momentum distribution supports this conclusion by indicating the presence of strong fluctuations and renormalizations 
in the metallic phase.  It suggests that some anomalous metal is stabilized at $U/t$ less than $4.75$.  This unusual metallic behavior is one of the most remarkable feature in our phase diagram.   

For both cases of $t'$, a nonmagnetic insulator phase is stabilized between the metal-insulator and magnetic transitions.  The antiferromagnetic long-range order is stabilized for the strong interaction regime in accordance with the expectation from  the corresponding $J_{1}$-$J_{2}$ Heisenberg model  derived from the strong coupling expansion in terms of $t/U$ and $t'/U$, while a spin-liquid insulator may be realized in this intermediate interaction regime.  According to the second-order perturbation in the limit of large interactions, the half-filled Hubbard model with the interaction $U$ and transfer $t_{ij}$ corresponds to the Heisenberg model with the superexchange coupling $J_{ij}=4|t_{ij}|^{2}/U$. Therefore the Hubbard model with $t'/t=0.5$ in the strong interaction limit  corresponds to the $J_{1}$-$J_{2}$ model with $J_{2}/J_{1}\simeq 0.25$, the ground state of which seems to have an antiferromagnetic long-range order of the N\'{e}el state.  Although we have seen that the commensurate antiferromagnetic long-range order appears in the Hubbard model with $t'/t=0.5,U/t> 7.25\pm 0.25$, the long-range order becomes absent for the weaker interaction.  We interpret it in the following way:  The double occupancy allowed at finite $U$ enhances the quantum fluctuation and reduce the magnetic local moment.  This charge fluctuation reduces the magnetic long-range order and antiferromagnetic order is destroyed before the transition to a metal.  The overall phase diagram also implies that the nonmagnetic insulator phase in the $J_{1}$-$J_{2}$ Heisenberg model is continuously connected with the nonmagnetic insulator phase in the present results, although we reserve the possibility that the dimer ordering speculated in the Heisenberg limit may be destroyed again by the charge fluctuation.   

Several discussions and comments on open issues may now be in order.  We note the possibility of observing the nonmagnetic insulating phase in real materials.  As we see in the phase diagram, Fig.~\ref{phdiag}, the width of this phase as a function of $U/t$ becomes very narrow for small $U_{c}$ and could easily become a concurrent transition  if the coupling to lattice degrees of freedom would be strong or effects of orbital degeneracy 
is present.  Several organic materials may belong to this class, although further careful analysis is desired.  It is also desired to explore possibility of materials which have larger effective $U$.
An interesting question to be examined experimentally is
whether the nonmagnetic insulator is generically stabilized near the 
metal-insulator transition point when the `frustration' effect such as $t'$
is large.
The nature of the unusual metal near the nonmagnetic insulator is an intriguing open question to be answered in future studies.  This is particulary interesting because the spin fluctuation is not the origin of the strong renormalization.  
  We have not studied the spin excitation gap. The presence or the absence of the spin excitation gap in the phase diagram is left for further studies.   
The nature of the nonmagnetic insulator state should be 
clarified in more detail in future.

   The bandwidth-control transitions in two dimensions clarified in this paper show sharp contrasts with the filling-control transitions.  The first clear difference is seen in the separation of the metal-insulator and antiferromagnetic transitions.  In the filling-control case, both transitions seem to occur simultaneously at the doping concentration $\delta =0$, if the antiferromagnetic order exists in the insulating phase.\cite{QMC2}  However, the two transitions occur at different points when the bandwidth is controlled.  In the filling-control transition, the antiferromagnetic short-ranged correlation critically grows in the paramagnetic metal with decreasing $\delta$ as $S(\pi,\pi)\propto\delta^{-1}$.  Such critical enhancement with continuous character of the transition is not clearly observed in the present bandwidth control and the first-order transition is plausible for small $t'$.  The metal-insulator transition is also clearly different each other:  The filling-control transition shows unusual but continuous decrease of the Drude weight\cite{I1998} indicating the continuous transition with 
specific scaling laws satisfied,\cite{fillcon} while the bandwidth control shows the level crossing with the first-order transition.  The proximity of the Mott insulator in the metallic phase with an anomalous metallic state is an outstanding feature of the filling-control transition while such fluctuations may not be expected for the  first-order transition inferred in the bandwidth control transition 
at smaller $U_c/t$.  However, it should also be noted that anomalous metallic states are suggested near the transition even in the bandwidth control as discussed above, if the transition occurs at large $U$ owing to large $t'/t$ 
with a continuous character of the transition as also suggested in the momentum distribution at $t'/t=0.5$.  

It would be an intriguing future problem to examine the generality of the 
present remarkable features of the bandwidth-control matal-insulator transitions
in other lattice structures or in three-dimensional systems.  The effect of 
orbital degeneracy is also an interesting subject to be studied.  
In real materials, there exist few examples of antiferromagnetic metals such as 
NiS$_{2-x}$Se$_x$ and V$_{2-y}$O$_{3}$, where the order of the transitions 
between the metal-insulator and antiferromagnetic transitions is opposite to the 
present case.  
From the present study, the existence of the antiferromagnetic metal
seems to be ascribed to a subtlety of the higher dimensionality and/or
presence of orbital degeneracy.  
                   
\section*{Acknowledgments} 
We thank S. Sachdev for useful discussions.  The work is supported by the `Research for the Future' program from the Japan Society
for the Promotion of Science under grant number JSPS-RFTF97P01103. A part of our computation has been done at the supercomputer center at the Institute for Solid 
State Physics, University of Tokyo.   
       

\begin{thebibliography}{999} 
\bibitem{MIT}For a review see M. Imada,A. Fujimori and Y. Tokura: Rev. Mod. Phys. {\bf 70} (1998) 1039. %
\bibitem{Mott}N. F. Mott and R. Peierls: Proc. Phys. Soc. London Ser. A{\bf 49} (1937) 72.  %
\bibitem{Slater}J. C. Slater:Phys. Rev. {\bf 82} (1951) 538. %
\bibitem{QMC1}M. Imada and Y. Hatsugai: J. Phys. Soc. Jpn. {\bf 58} (1989) 3752. %
\bibitem{QMC2}N. Furukawa and M. Imada: J. Phys. Soc. Jpn. {\bf 61} (1992) 3331. %
\bibitem{Assaad}F. F. Assaad and M. Imada: Phys. Rev. Lett. {\bf 76} (1976) 3176. %
\bibitem{PIRG}M. Imada and T. Kashima: J. Phys. Soc. Jpn. {\bf 69} (2000) 2723.%
\bibitem{oldHub}J. Hubbard: Proc. Roy. Soc. London.A{\bf 276} (1963) 238;A{\bf 277} (1964) 237;A{\bf 281} (1964) 401 %
\bibitem{KANA}J. Kanamori:  Prog. Theor. Phys.{\bf 30}(1963)275. %
\bibitem{EXTRAPO}T. Kashima and M. Imada: cond-mat/0104140 and unpublished. %
\bibitem{fillcon}M. Imada: J. Phys. Soc. Jpn. {\bf 64} (1995) 2954. %
\bibitem{tprime1}M. S. Hybertsen, M. Schl\"{u}ter and N. E. Christensen: Phys. Rev. B{\bf 39} (1989) 9028 %
\bibitem{tprime2}A. K. McMahan,R. M. Martin and S. Satpathy: Phys. Rev. B{\bf 38} (1988) 6650 %
\bibitem{tprime}K. Okada and A. Kotani:  J. Phys. Soc. Jpn. {\bf 58} (1989) 1095 %
\bibitem{dp_H1}M. S. Hybertsen, E. B. Stechel, M. Schl\"{u}ter and D. R. Jennison:  Phys.Rev.B{\bf 41} (1990) 11068 %
\bibitem{Andersen}For example, see E. Pavarini, I. Dasgupta, T. Saha-Dasgupta, O. Jepsen and O.K. Andersen: cond-mat/0012051. %
\bibitem{HF}H. Kondo and T. Moriya: J. Phys. Soc. Jpn. {\bf 65} (1996) 2559 %
\bibitem{HF2}H. Kondo and T. Moriya: J. Phys. Soc. Jpn. {\bf 67} (1998) 234 %
\bibitem{GA1}T. Ogawa,K. Kanda and T. Matsubara: Prog. Theor. Phys. {\bf 53} (1975) 614 %
\bibitem{GA2}F. Takano and M. Uchinami: Prog. Theor. Phys. {\bf 53} (1975) 1267 %
\bibitem{VMC1}H. Yokoyama and H. Shiba: J. Phys. Soc. Jpn. {\bf 56} (1987) 1490 %
\bibitem{VMC2}H. Yokoyama and H. Shiba: J. Phys. Soc. Jpn. {\bf 56} (1987) 3582 %
\bibitem{Hirsch}H. Q. Lin and J. E. Hirsch: Phys. Rev. B{\bf 35} (1987) 3359 %
\bibitem{firstliquid}P. Chandra and B. Doucot: Phys. Rev. B{\bf 38} (1988) 9335 %
\bibitem{Gelfand}M.P. Gelfand, R. R. P. Singh, and D. A. Huse:
Phys. Rev. B{\bf 40} (1989) 10801 
%
\bibitem{Nakamura}T. Nakamura, and N. Hatano:
J. Phys. Soc. Jpn.{\bf 62} (1993) 3062.
%
\bibitem{refspinliquid}A. V. Dotsenko and O. P. Sushkov: Phys. Rev. B{\bf 50} (1994) 13821 %
\bibitem{SL_CO}J. Oitmaa and Z. Weihong: Phys. Rev. B{\bf 54} (1996) 3022 %
\bibitem{CO}S. Miyazawa and S. Homma: Phys. Lett.A{\bf 193} (1994) 370 %
\bibitem{Sachdev}N. Read and S. Sachdev: Phys. Rev. Lett. {\bf 62} (1989) 1694 %
\bibitem{Sushkov}O. P. Sushkov, J. Oitmaa and Z. Weihong: Phys. Rev. B{\bf 63} (2001) 104420 %
\bibitem{Sorella2}L. Capriotti and S. Sorella, : Phys. Rev. Lett. {\bf 84} (2000) 3173 %
\bibitem{I1998}H. Tsunetsugu and M. Imada: J. Phys. Soc. Jpn. {\bf 67} (1998) 1864 
\end{thebibliography}
\end{document}